\documentclass[fleqn,usenatbib]{mnras}

\usepackage{newtxtext,newtxmath}
\usepackage[T1]{fontenc}
\usepackage{ae,aecompl}

\usepackage{graphicx}	% Including figure files
\usepackage{amsmath}	% Advanced maths commands
\usepackage{amssymb}	% Extra maths symbols
\usepackage{subcaption}
\usepackage[percent]{overpic}
\usepackage{verbatim} % adds environment for commenting out blocks of text & for
\usepackage{comment}
\usepackage{tikz}
\usepackage{adjustbox}

\title[]{Strong lensing reveals jets in a sub-microJy radio quiet quasar}
\author[P. Hartley et al.]{
P. Hartley,$^{1}$\thanks{E-mail: philippahartley@hotmail.com}
N. Jackson,$^{1}$
D. Sluse,$^{2}$
H. R. Stacey$^{3,4}$
and H. Vives-Arias$^{5,6}$
\\
$^{1}$Jodrell Bank Centre for Astrophysics, School of Physics \& Astronomy, University of Manchester, Oxford Rd., Manchester M13 9PL, UK\\
$^{2}$ STAR Institute, Quartier Agora - All\'ee du six Ao\^ut, 19c B-4000 Li\`ege, Belgium\\
$^{3}$ ASTRON, Netherlands Institute for Radio Astronomy, Oude Hoogeveensedijk 4, 7991 PD, Dwingeloo, The Netherlands\\
$^{4}$ Kapteyn Astronomical Institute, PO Box 800, 9700 AV Groningen, The Netherlands\\
$^{5}$Instituto de Astrof\'isica de Canarias, V\'ia L\'actea, s/n, E-38205 La Laguna, Tenerife, Spain\\
$^{6}$Departmento de Astrof\'isica, Universidad de La Laguna, E-38205, La Laguna, Tenerife, Spain
}

\date{Accepted XXX. Received YYY; in original form ZZZ}

\pubyear{2019}

\begin{document}
\label{firstpage}
\pagerange{\pageref{firstpage}--\pageref{lastpage}}
\maketitle

\begin{abstract}
We present e-MERLIN and EVN observations which reveal unambiguous jet activity within radio quiet quasar HS~0810+2554. With an intrinsic flux density of 880~nJy, this is the faintest radio source ever imaged. The findings present new evidence against the idea that radio loud and radio quiet quasars are powered by different underlying radio emission mechanisms, showing instead that the same AGN mechanism can operate as the dominant source of radio emission even in the very lowest radio luminosity quasars. Thanks to strong gravitational lensing, our source is not only visible, but with VLBI is imaged to a scale of just 0.27~pc: the highest ever resolution image of a radio quiet quasar. Brightness temperatures of at least $8.4\times 10^6$~K are associated with two highly compact components. Subsequent modelling of the lensed system has revealed that the components are linearly aligned on opposing sides of the optical quasar core, with the typical morphology of a compact symmetric object (CSO). Given that this source has been found to fall on the radio--FIR correlation,  we suggest that the radio--FIR correlation cannot always be used to rule out AGN activity in favour of star-formation activity. The correlation -- or at least its scatter -- may conceal the coexistence of kinetic and radiative feedback modes in AGN. Modelling of the lensing mass itself points to a non-smooth mass distribution, hinting at the presence of dark matter substructure which has manifested as astrometric perturbations of the VLBI lensed images.%, posing no threat to the CDM paradigm

\end{abstract}


%%%%%%%%%%%%%%%%%%%%%%%%%%%%%%%%%%%%%%%%%%%%%%%%%%

%%%%%%%%%%%%%%%%% BODY OF PAPER %%%%%%%%%%%%%%%%%%

\section{Introduction}

The strong gravitational lensing of a background object results in the magnification of its size and, since surface brightness is conserved, its observed flux density (see \citealt{2002LNP...608....1C,2006glsw.conf.....M,bart2010} for extensive lensing reviews and formalism). This powerful property allows us to use strong lenses as cosmic telescopes, typically increasing the effective sensitivity of the observing instrument to that of one 25--100 times bigger. With such a configuration we can make images of radio sources with nJy flux densities in a way that will only become routine with the Square Kilometre Array \citep{2015aska.confE..87O}. We can study these sources in more detail, resolving beam-limited structure into individual components (e.g. \citealt{2006A&A...451..865C,2011ApJ...739L..28J}).  Strong lensing also results in the multiple imaging of the background source, with typical image configurations consisting of either two or four magnified images in addition to a final, de-magnified, image. Since the geometry of the resulting pattern is dependent only on the distribution of matter between the source and the observer, careful modelling of observations of strong lens systems can reveal the nature and amount of this matter, both baryonic and dark. At large scales, modelling of image separations allows us to measure the lensing galaxy profile (see e.g. \citealt{2002ApJ...575...87T,2003MNRAS.343L..29T,2004ApJ...611..739T,0004-637X-752-2-163,0004-637X-777-2-98,0004-637X-800-2-94}).  At smaller scales, flux anomalies and image distortions allow us to probe galaxy substructure to look for evidence of dark matter sub-haloes  (see e.g. \citealt{1998MNRAS.295..587M,2001ApJ...563....9M,2004ApJ...610...69K,2012Natur.481..341V,2017MNRAS.471.2224N}). With the exploitation of these remarkable science applications in mind, investigations have recently been conducted by \cite{2015MNRAS.454..287J}, identifying the population of strongly-lensed radio quiet quasars  (RQQs) as useful candidates for observation due to the sensitivity of their sub-mas brightness structure to small-scale gravitational perturbations, their relative insensitivity to microlensing effects, and their significant contribution to the number of lensed objects at the faint end of the radio luminosity function.

Although originally discovered in a radio survey \citep{1963Natur.197.1040S}, most quasars actually produce relatively low levels of radio emission. The classic radio-loud quasar (RLQ) is typically observed to emit radio light from dramatic jets produced via the accretion of galactic gas and dust onto a central, supermassive, black hole. RQQs, on the other hand, are not associated with such obvious radio structure.  The distinction between RLQs and RQQs is conventionally made using the ratio $R$ of 5~GHz and optical B-band monochromatic luminosities, the cut being made at $R$=10. While early optically-selected samples found that most RQQs reside below $R$=1 \citep{1989AJ.....98.1195K}, source count studies undertaken by \cite{MNR:MNR1752} and \cite{2006A&A...455..161L} have found a less firm distinction between the two regimes, finding sources of intermediate flux density.

It is not known whether the radio emission that we do see in RQQs is due to the same active galactic nucleus (AGN) engine observed in RLQs, or whether starburst activity or other, more exotic, mechanisms are responsible. The possibilities have been debated extensively in the literature over several decades.  Work by \cite{0004-637X-768-1-37} used a sample from the NVSS to show that there is an upturn in source counts at the faint end of the quasar luminosity function, hinting at a large population of star-forming sources residing at the \textmu Jy level.  A simple ``unified'' model based on AGN orientation has been ruled out by \cite{2016arXiv160804586K}, and some authors cite a lack of jet features as evidence for a lack of radio AGN activity \citep{2017NatAs...1E.194P}. \cite{2016A&A...589L...2H}, however, have shown by using VLBA observations that the radio emission of at least some RQQs is dominated by an AGN. Others point to the apparent flux density continuum, along with evidence of source variability by \cite{0004-637X-618-1-108}, to argue for a model where the same AGN engine is operating -- at differing levels of power -- in each case. Alternative approaches to the problem have made use of the tight radio--far-infrared (radio--FIR) flux density correlation observed in star-forming regions \citep{doi:10.1093/mnras/251.1.14P}. \cite{2017MNRAS.468..217W} showed an excess of radio emission compared to that expected from the radio--FIR correlation in a sample of QSOs, which cannot be explained by star formation alone. This result contradicted the findings of \cite{2013MNRAS.436.3759B,2015MNRAS.453.1079B}. Optically-thin bremsstrahlung emission in the core of the sources \citep{2007ApJ...668L.103B} and exotic mechanisms such as emission from magnetically-heated coronae \citep{MNR:MNR13806, 2018arXiv181010245L} and radiatively-driven shock fronts \citep{2014MNRAS.442..784Z} are also suggested. 

By determining the RQQ radio emission process we will be able to address the wider problem of galaxy evolution, in particular the role of AGN feedback and whether AGN activity causes suppression or, alternatively, activation of star formation. Suspicion that the two processes are linked has arisen from evidence of corresponding trajectories over cosmic history \citep{2014ARA&A..52..415M}. However, it remains possible that the apparent connection is due simply to the same availability of gas over time. Many previous investigations have been statistical in nature and are subject to potential biases arising from sample selection. Additionally, some investigations have relied on the use of assumed spectral indices when converting catalogues of observed flux densities from one waveband to another. The use of observations of high resolution and sensitivity, in combination with the magnifying power of lensing, allows us to avoid these problems and instead determine the nature of the radio structure by directly imaging it.

Radio observations of lensed quasars not only provide the magnification needed to expand our knowledge of the RQQ population, but also lend ideal conditions for the study of dark matter structure (see \cite{2010AdAst2010E...9Z,2013BASI...41...19J} for recent reviews). The `missing satellite problem' describing the paucity of evidence for sub-structure \citep{1999ApJ...524L..19M,1999ApJ...522...82K} persists, despite recent observations that have found some ultra-faint dwarf galaxies in proximity to our own \citep{2006ApJ...647L.111B,2006ApJ...643L.103Z,2015ApJ...805..130K,2016ApJ...832...21H} and a halving of the overall halo mass function predicted by the Illustris simutations \citep{2017arXiv171011148G}. While observation of lensed extended sources can probe the higher-mass end of the substructure mass function ($10^9$~${\rm M}_{\odot}$: \citealt{2003ApJ...590..673W,0004-637X-623-1-31,2005MNRAS.363.1136K,2010MNRAS.408.1969V,2012Natur.481..341V,2018arXiv181103627R}), quasar lens systems provide a way of probing scales down to $10^6 {\rm M}_{\odot}$ \citep{1998MNRAS.295..587M}. 

Flux density perturbations due to small-scale structure, which can be betrayed by flux anomalies between merging images in cusp and fold configurations, have been observed in several lensed systems (see e.g.  \citealt{MNR:MNR19729,0004-637X-773-1-35}; Badole et al. 2018, in preparation).  Radio observations of lensed quasars are particularly suitable for this study since, in the same manner as narrow-line regions \citep{2017MNRAS.471.2224N}, microlensing events are likely to be rare. Unlike in the optical, where microlensing events occur frequently, only a single radio microlensing event has been observed to date \citep{2001ASPC..239..363K}. It is possible, however, that any flux density anomalies found in radio observations of lensed quasars could also be due to intrinsic AGN variability, to scattering by galactic scintillation or intervening plasma clouds (see e.g., \citealt{phillips319,0004-637X-595-2-712,2004evn..conf..163B}), to the differential magnification of an extended background source \citep{2012MNRAS.424.2429S}, or to disk and other baryonic structures within the lensing galaxy \citep{doi:10.1046/j.1365-8711.1998.01304.x,2003yCat..73450001M,2015MNRAS.447.3189X,2018MNRAS.475.2438H}. Astrometric perturbations of lensed images are harder to detect; several studies  \citep{0004-637X-659-1-52,2001ApJ...563....9M,2013MNRAS.431.2172Z} predict offsets of just $\sim$1--10 mas between lens models containing a smooth mass distribution and those containing dark matter substructure. Such small perturbations can be revealed by using the very high resolution ($\sim$1 mas) observations of Very Long Baseline Interferometry (VLBI), and attempts by \cite{2004MNRAS.350..949B} and \cite{2018MNRAS.478.4816S} to fit smooth models to VLBI quasar observations have indeed struggled to obtain a satisfactory fit, strongly suggesting the presence of substructure in these examples.  \cite{refId0} have found, however, that small galaxies located within a few arcseconds of a lens but unaccounted for explicitly in the model can also produce astrometric anomalies of the order of a few mas.

In this paper we expand on the work of \cite{2015MNRAS.454..287J}, who recently used the Karl G. Jansky Very Large Array (JVLA) and the extended Multi Element Remote Linked Interferometer Network (e-MERLIN) to detect a total of six strongly lensed RQQs. The observations implied an intrinsic lensed source flux density of just 1--5 \textmu Jy, constituting some of the very faintest radio sources ever detected. Two sources were subsequently detected at higher resolution using the e-MERLIN array.  From this sample, we have selected QSO HS~0810+2554 for further investigation. Discovered by \cite{2002A&A...382L..26R}, HS~0810+2554 is a fold-configuration quad lens system with the source, a narrow-line quasar, at redshift $z = 1.51$ and a main lensing galaxy at an approximate redshift of 0.89 \citep{2011ApJ...738...96M}. \cite{2015MNRAS.454..287J} have measured a spectral index for the radio emission of $-0.55\pm 0.1$, and have modelled the system to predict intrinsic sizes of the background radio source of $<$2~kpc and extended linearly. This scale, smaller than that typically observed in star-forming regions \citep{2012MNRAS.422.3339W,2005MNRAS.358.1159M}, strongly suggests the presence of an AGN mechanism at play, producing radio emission from compact jet and core components. In the case of HS~0810+2554, the flux anomaly recorded in the optical was not apparent in the radio observations conducted by \cite{2015MNRAS.454..287J}. For our study we have used the European VLBI Network (EVN) to investigate whether the compact structure predicted is in the jet/core form of an AGN or in the form of relatively diffuse starburst emission, invisible on the spatial scale of EVN baselines. We used the resulting radio map to model the lens mass distribution and reconstruct the background source. We present imaging and modelling results from this observation, alongside further observations made using e-MERLIN.

\section{Observations and calibration strategy}

\subsection{EVN observation}

Radio continuum observations of HS 0810+2554 were carried out on March 3rd and 4th 2016 using the EVN (projects EJ016A and EJ016B) at 1.6 GHz, over a frequency range of 1539--1667 MHz. The total bandwidth was divided into 8 adjacent spectral windows, each divided into 32 channels and a visibility integration time of 2~s was used. A total of 14 telescope dishes were combined into a synthesised aperture, spanning baseline lengths up to 7000 km to achieve a beam-limited resolution of the order 0.001 arcsec. A nearby point source, J~0813+2435, was observed in order to obtain phase corrections.  4C39.25 was observed in order to adjust for the bandpass. The largest dish, Effelsberg, was only available for one of the two days of observation.

Data were initially reduced using the EVN pipeline, which uses a {\sc parseltongue} wrapper to perform calibration and imaging routines from the Astronomical Image Processing System ({\sc aips}) software package distributed by NRAO \citep{greisson}. Fringe fitting was performed to remove instrumental and atmospheric delay, using observations of the phase calibrator and bandpass calibrator. Inspection of delay solutions revealed long delays induced at the 76m Lovell telescope at Jodrell Bank during the first epoch, which, due to long intervals between phase referencing necessitated by long slew times, resulted in an incorrect interpolation of delay solutions for baselines to this antenna. Data containing the Lovell was therefore lost for half of the total observation time. Phase and amplitude calibration was also performed using the phase calibrator with a point source model, and the solutions transferred to the target source. Pipeline solutions were refined by carrying out additional flagging of radio-frequency interference (RFI) and by performing additional manual self-calibration on the phase calibrator. Self-calibration of the target source was attempted but solutions were unstable due to a small number of larger antennas dominating the signal, despite experiments which downweighted these stations. Final maps were produced using three different weighting strategies, ranging from fully natural weighting, which maximises sensitivity, to a compromise weighting between natural and uniform, which modifies the weights of individual visibilities by their local density in the $u-v$ plane, upweighting longer-baselines to increase resolution. 

\subsection{e-MERLIN observations}

Radio observations of HS~0810+2554 were carried out using the e-MERLIN array at about 5~GHz (C~band). The 76m Lovell dish -- the largest of the array -- was not available for this study. Three 24 hour observations were performed, each at a different 512~MHz bandwidth within the tunable C~band range of 4-8~GHz. Each 24 hour observation was performed across three epochs, resulting in a total of nine visibility sets for calibration and imaging. Observations on 14th and 15th December 2015 and 28th Feb 2016 used a frequency range centred on 5.12~GHz, 12th, 13th and 14th May 2016 used 4.32~GHz and 15th, 16th and 17th May 2016 used $7.12$~GHz. Each epoch used an integration time of 1 second and the total bandwidth was divided into 4 IFs, each further divided by a channelisation of 512, resulting in channels 0.25~MHz wide. Phase referencing was performed using the nearby bright point source J~0813+2435, 3C286 was observed in order to set the flux scale, and the bright point source OQ208 was used to calculate the bandpass response.

Data reduction followed standard procedure \citep{2015arXiv150204936A}, again using the {\sc aips} package. Initial inspection of the visibility data showed significant RFI contributions and antenna dropouts variously affected different antennas and calibrator sources. Additionally, the Defford antenna -- known to have low sensitivity at C~band -- showed little signal over the background noise during the last six epochs, with too much scatter to determine an accurate phase slope in the frequency domain. This data from this telescope was therefore discarded from the $4.32$~GHz and $7.32$~GHz observations. After performing nominal delay subtraction by fringe-fitting the data from all calibration sources, separate complex gain calibration was performed, followed by bandpass calibration and the setting of the overall flux scales. Gain solutions for the phase reference source were refined using self-calibration with a point source model, and solutions transferred to the the target. The described procedure was performed individually for each epoch, before the visibility data of epochs of matching frequency were combined for imaging and further calibration.

Self-calibration of the faint target was aided by the presence of a $\sim$200 \textmu Jy confusing source located close to the edge of the usable field of view (F.o.V.), 6 arcmin from the pointing centre. Using a point source model for this source resulted in a significant refinement of residual phase errors arising from atmospheric instabilities in the vicinity of the target, improving the dynamic range considerably in the 5.1~GHz and 4.32~GHz maps. The technique relied upon the use of unaveraged visibility data in order to avoid significant smearing of the confusing source. Unfortunately, smearing resulting in $\sim$30\% intensity loss could not be avoided in the 7.32~GHz map, and self-calibration was not successful in this case. Final maps were produced using natural weighting.

\section{Radio maps}

\subsection{EVN map}

\begin{figure*}
  \centering
      \includegraphics[trim={3.8cm 0cm 5cm 3cm},clip,width=1\linewidth]{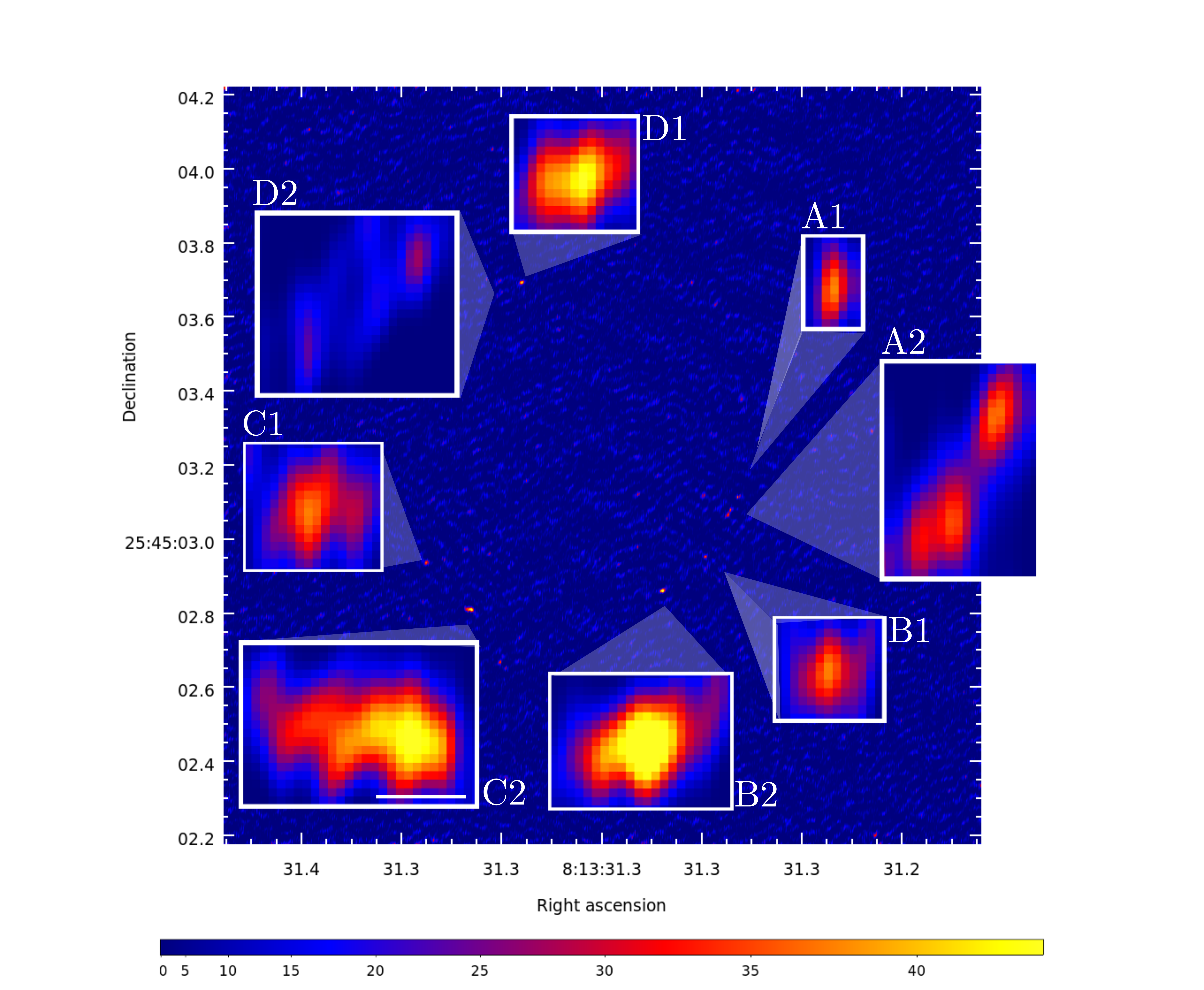} 
\caption{EVN image of HS~0810+2554 at 1.65~GHz produced using a natural weighting scheme.  The peak surface brightness is 52~\textmu Jy beam$^{-1}$, at component B2. The beam is at full width of half maximum (FWHM) at 12.0 $\times$ 8.5~mas at a position angle of -3$^{\circ}$. We use the nomenclature of \protect\cite{2002A&A...382L..26R} for respective pairs of components associated with background sources 1 and 2. A contour plot of the uncovered background is supplied in~\ref{contall}.}
 \label{0810evncomposite}
\end{figure*}

The naturally weighted EVN image of 0810+2554 is shown in Fig.\ref{0810evncomposite}. The off-source r.m.s. noise level is 7~\textmu Jy. The image shows a clear detection of the lensed RQQ, with several components visible over $5\sigma_{\rm r.m.s.}$. A total of eight lensed components are identified, indicating that the EVN has resolved the extended background structure observed by e-MERLIN at L~band into two separate components, here referred to as source 1 and source 2. In order to correctly associate components of the lensed image with components of the background source, the modelling process discussed in Section~\ref{model} was iterated over all combinations of four pairs within the total of eight lensed components. This revealed a clear preference for the configuration notated in Fig.\ref{0810evncomposite}, with a five-fold improvement in goodness-of-fit over the second-preferred configuration. 

%Further confirmation was found by comparing relative astrometric positions  the optical HST positions measured by ... The HST positions of merging images A and B are approximately coincident with the positions A1 and B1, indicating that source 1 is situated very close to the quasar core. 

\begin{figure*}

      \begin{subfigure}[h]{1\textwidth}
        \caption[]{Components imaged using a restoring beam FWHM of 12$\times$8.5~mas. }
  \centering
  \includegraphics[trim={0cm 17cm 0cm 2.5cm},clip,width=0.8\columnwidth]{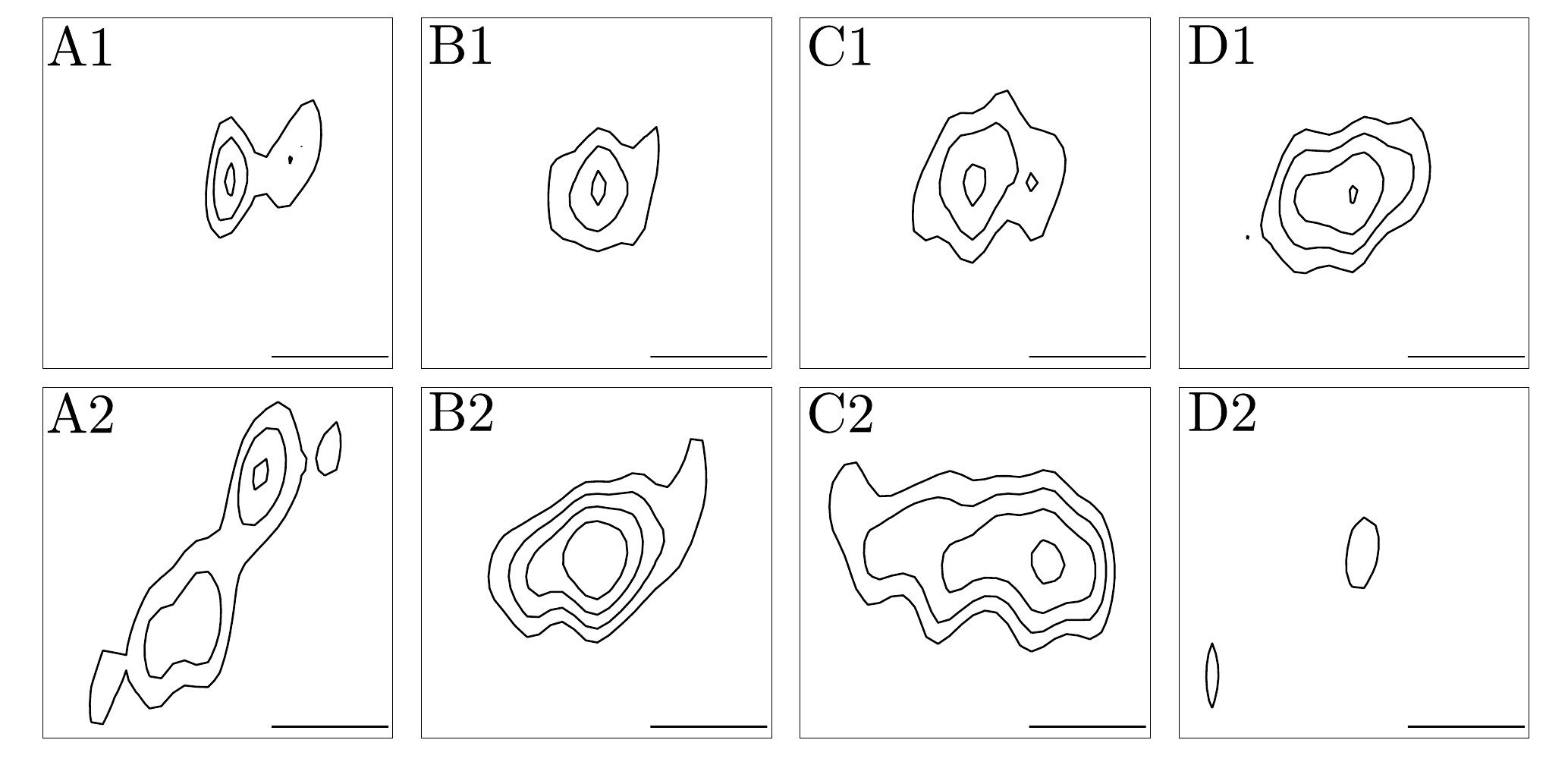}
  \label{inspectspace}
  \end{subfigure}

        \begin{subfigure}[h]{1\textwidth}
          \caption[]{Components imaged using a restoring beam FWHM of 12.8$\times$4.0~mas. }
  \centering
  \includegraphics[trim={0cm 7cm 0cm 12.7cm},clip,width=0.8\columnwidth]{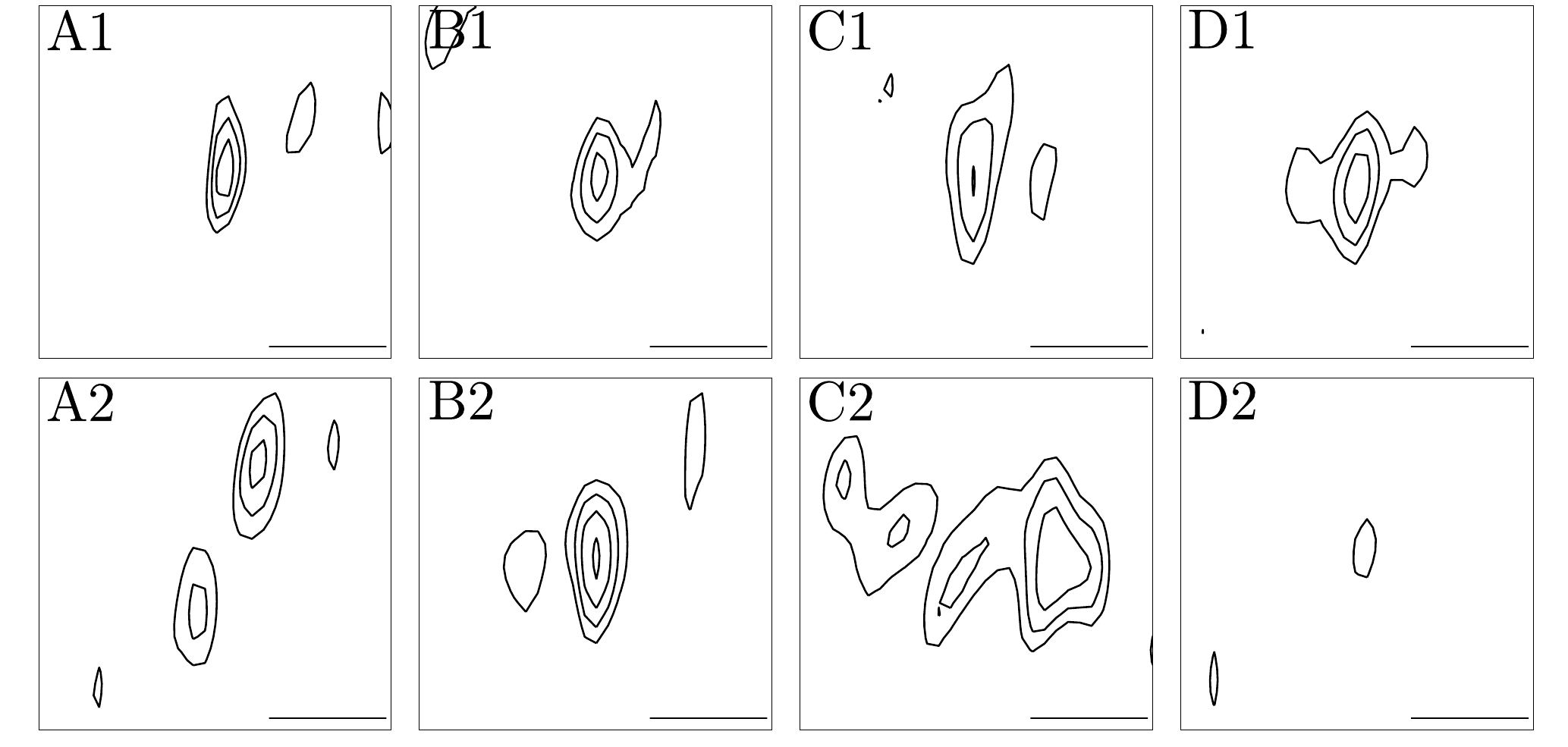} 

  \label{inspectspace}
  \end{subfigure}
  
        \begin{subfigure}[h]{1\textwidth}
  \centering
    \caption[]{Components imaged using a restoring beam FWHM of 11.1$\times$3.4~mas}
  \includegraphics[trim={0cm 17cm 0cm 2.5cm},clip,width=0.8\columnwidth]{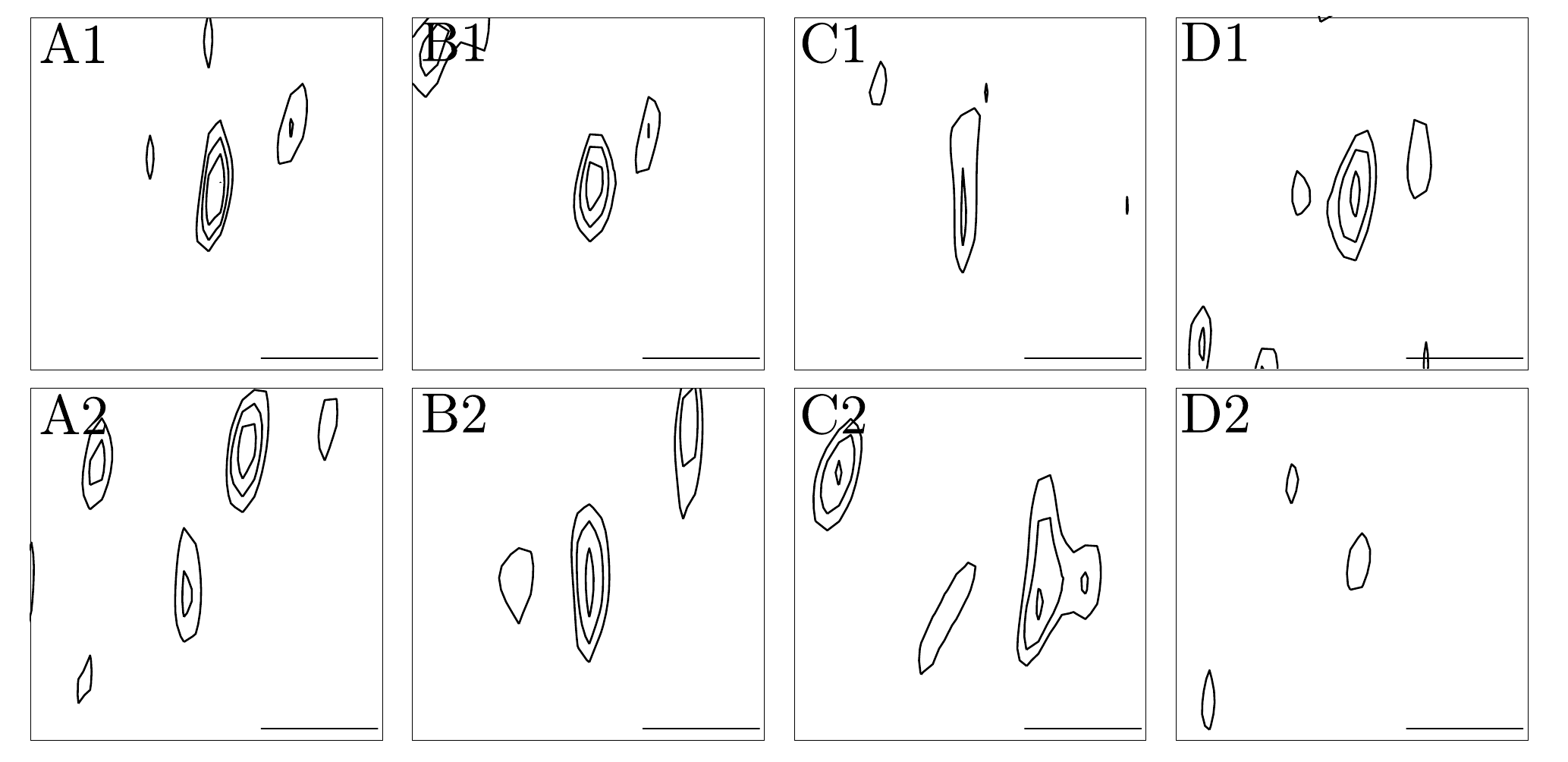} 

  \label{inspectspace}
  \end{subfigure}

    \caption[]{Maps of the lensed components of HS~0810+2554 made using three different weighting schemes, from a natural weighting scheme (top) to a compromise weighting scheme between natural and uniform (bottom), with robustness parameters of 5, 1 and 0 in the Briggs' weighting scheme, respectively. In all images contours have been drawn at (-3, 3, 4, 5, 6)~$\times$~7.2~\textmu Jy and the scalebar represents a length of 10 mas.}
    \label{morphdetail}
 \end{figure*}

The maps produced using the two mixed weighting schemes provide additional insight into the nature of the two background sources. Fig.~\ref{morphdetail} displays each component image using restoring beam sizes of 12.0$\times$8.5~mas, 12.8$\times$4.0~mas and 11.1$\times$3.4~mas for weighting schemes which used a Briggs' robustness parameter of 5, 1 and 0, respectively. As increasingly higher resolution is used, components A2, B2 and C2 become distinctly resolved into two separate unresolved sub-components. In the map using the highest resolution, all components in the map are completely unresolved, suggesting that the two background sources are both highly compact. Flux density measurements for all components at all resolutions were made by using {\sc aips} task {\sc jmfit}, which fits Gaussian components to the sky brightness distribution. A double Gaussian model was used to fit the flux density components belonging to source 2, and a single Gaussian used for source 1. The results, along with associated errors, are reported in Table~\ref{0810robust5}.

The flux density measurements were used in order to place a lower limit of the brightness temperatures of the radio components in the source plane. According to the Rayleigh Jeans law, the brightness temperature, $T_{\rm B}$, of a non-thermal source can be written: 

%\begin{equation}
%T = \frac{\lambda^2}{2k\Omega}S,
%\end{equation}

\begin{equation}
T_{\rm B} = 1222 \times \frac{I}{\nu^2 \theta_{\rm maj} \theta_{\rm min}},
\end{equation}

%MKS (SI) units
where $\nu$ is the observing frequency in GHz, $\theta_{\rm maj}$ and $\theta_{\rm min}$ are the half-power beam widths along the major and minor axes, respectively, measured in arcseconds, and $I$ is the source brightness intensity measured in mJy beam$^{-1}$. At an observing frequency of  $1.65$~GHz,  and with a restoring beam of $\theta_{\rm maj} =11.1$ and $\theta_{\rm min} = 3.4$ mas, the measurements of flux density per beam from Table~\ref{0810robust5} give maximum values of $0.045\pm 0.009$~mJy and $0.044\pm 0.009$~mJy for source 1 and source 2, respectively. According to {\sc aips} task {\sc jmfit}, the sources are unresolved in all lensed components at this resolution. Since a source is considered to be unresolved if its deconvolved FWHM size is smaller than one-half the synthesised beam FWHM, and since the lensed components are magnified in orthogonal directions, we assume an upper limit to the physical source size of $\theta_{\rm maj}=1.55$ mas and $\theta_{\rm min}=1.55$ mas, using half the value of the beam size as is conventional for radio-continuum observations (e.g. \citealt{mundell2000parsec,ulvestad2001origin,2017A&A...597A...5M}). Brightness temperatures of $8.4\:(\pm0.3)\times 10^6$~K and $8.2\:(\pm0.3)\times10^6$~K therefore represent lower limits. Using the spectral index determined using our e-MERLIN observation below, the K-corrected flux densities in the $z = 1.51$ rest frame of the source yield brightness temperatures of at least $6.0\:(\pm0.6)\times10^6$~K and $5.8\:(\pm0.5)\times 10^6$~K.

%, with nominal major and minor axis sizes of 0.0 after deconvolving the components from the clean beam

 \begin{table}
	\centering
	\begin{tabular}{cccc} % four columns, alignment for each
    \hline
    	\hline \noalign{\smallskip}
	     Component &12$\times$8.5~mas&12.8$\times$4.0~mas&11.1$\times$3.4~mas\\
	      		\hline \noalign{\smallskip}
	    A1&$37\pm 7$&$39\pm 8$ &$44\pm9$\\
	    A2a&$35\pm 7  $ &$38\pm 8$&$37\pm 8$\\
        A2b&$32\pm 7  $ &$27\pm 8$&$24\pm 8$\\
	    B1& $32\pm 7$ &$35 \pm 9$&$37 \pm 8$\\
	    B2a& $52\pm 7$  &$39 \pm 8$&$45\pm9$\\
    	B2b& $41\pm 7$  &$21 \pm 8$&$29\pm9$\\
	    C1&$35\pm 7$ &$35\pm 8$&$35\pm 8$\\
	    C2a&$45\pm 7$&$41 \pm 8$ &$41\pm 8$\\
	    C2b&$34\pm 7$&$29 \pm 8$ &$31\pm 8$\\
	    D1&$43\pm 7$ &$46\pm 8$&$41\pm9$\\
	    D2a&  $20 \pm 7$ &$25\pm 9  $&$27 \pm 9$\\
        D2b&  $20 \pm 7$ &$21\pm 8 $&$ 22 \pm 9$\\

	%	\\
		\noalign{\smallskip}\hline
        \hline
	\end{tabular}
    	\caption{Radio flux density measurements made in \textmu Jy of the components of HS~0810+2554 observed with the EVN at 1.65~GHz. Measurements of each component were taken from images made using three different weighting schemes, from a natural weighting scheme to a compromise weighting scheme between natural and uniform, with robustness parameters of 5, 1 and 0 in the Briggs' weighting scheme, respectively. The values corresponding to source 2 were measured by fitting two Gaussian components, a and b, in each case. Errors are derived from {\sc aips} task {\sc jmfit} and do not include an uncertainty on the setting of the overall flux density scale 5--10\%.}
        \label{0810robust5}
\end{table}

\subsection{e-MERLIN maps}

\begin{figure*}
  \centering
  \begin{tabular}{cc}
  \includegraphics[trim={1.4cm 4.3cm 1.6cm 5.5cm},clip,width=1\columnwidth]{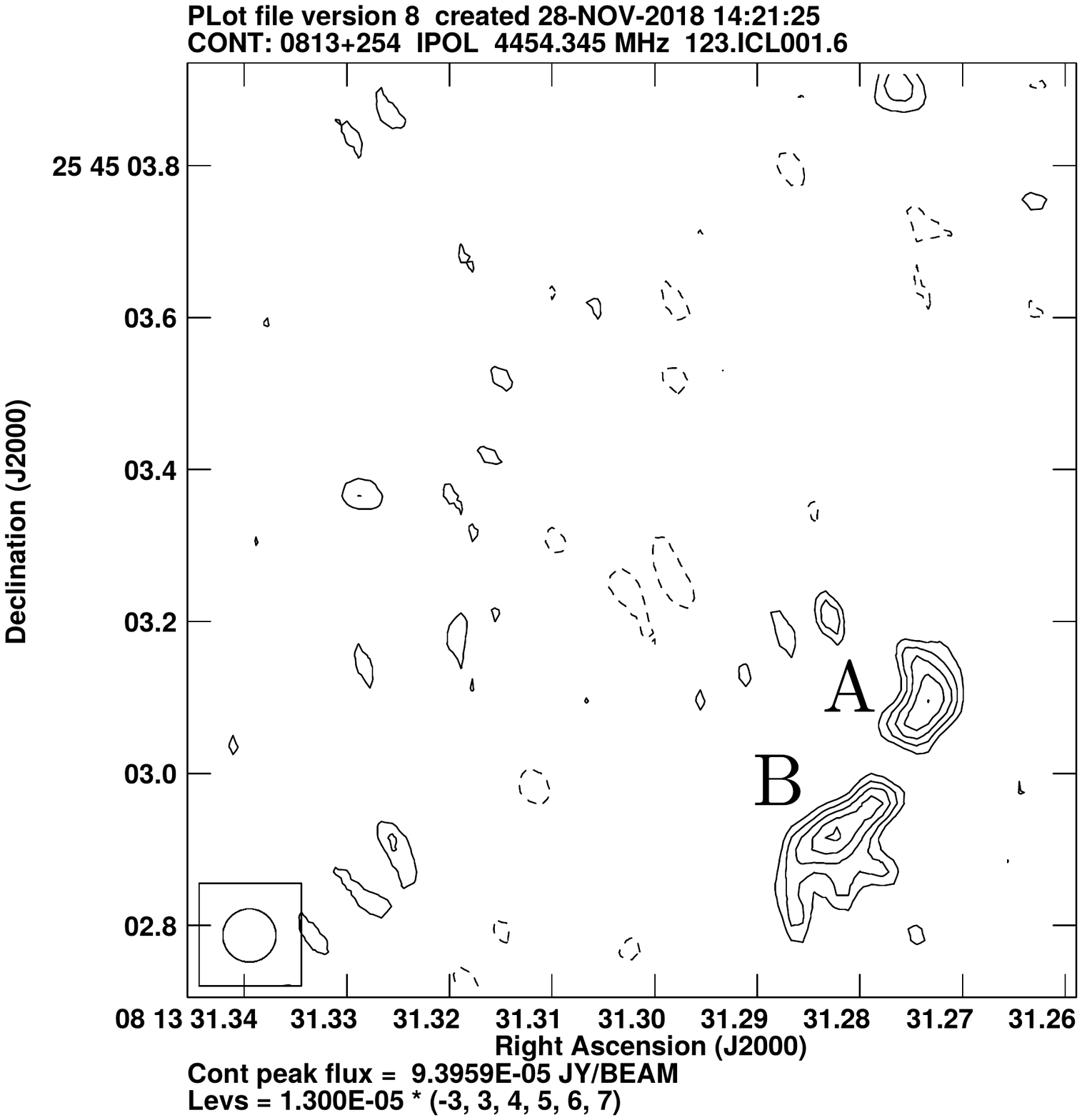}
  &
  \includegraphics[trim={1.4cm 4.3cm 1.6cm 5.5cm},clip,width=1\columnwidth]{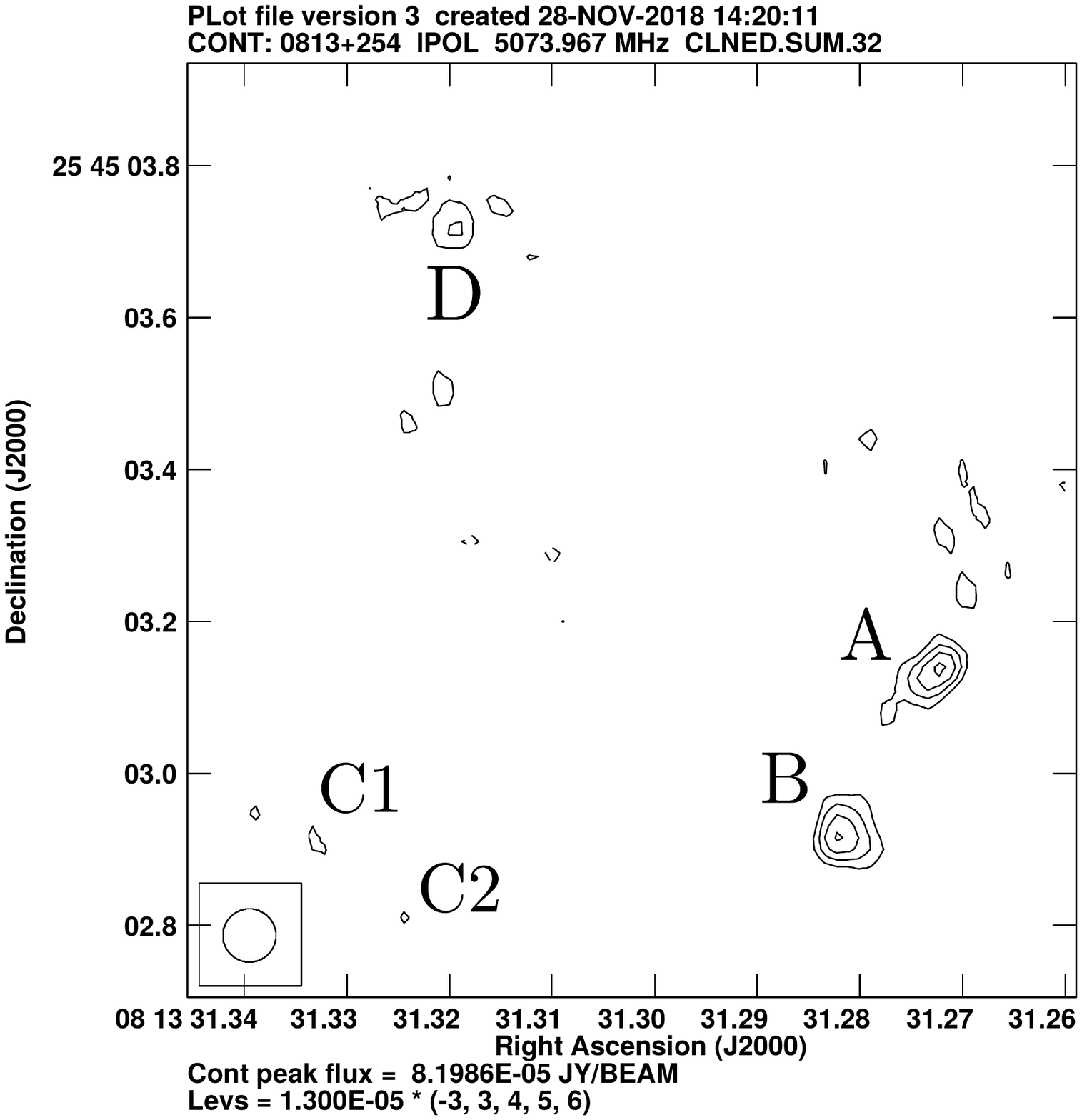}
\\
  \end{tabular}
  \caption{Maps of HS~0810+2554 made from e-MERLIN observations at 4.32~GHz (left) and 5.12~GHz (right). Both maps were produced using the same 70 $\times$ 70 mas restoring beam. Both maps also use the same contouring scheme, at values of (-3, 3, 4, 5, 6, 7) $\times$ 13 \textmu Jy, which is the value of the off-source r.m.s. noise in the higher-sensitivity 5.12~GHz map. Image components with signal higher than the respective value of 3$\sigma_{\rm r.m.s.}$ in each map are labelled.}
  \label{123456}
 \end{figure*}

Final e-MERLIN maps observed at 4.32~GHz and 5.12~GHz are shown in Fig.~\ref{123456}. The 7.32~GHz map, which suffered both from the loss of the Defford station and from the lack of self-calibration on the target, showed no detection of the source. In order to allow direct comparison between the two frequencies, both maps were produced using the same circular restoring beam of 70 $\times$ 70 mas. In the 5.12~GHz map component~C is resolved into the two sub-components seen in the EVN map, but is not prominent within the higher noise of the 4.32~GHz map. Flux density measurements for all components in each map were again obtained using {\sc jmfit}. The results are presented in Table~\ref{emerlintargetfluxes}. There is no evidence of a flux anomaly between images A and B in either map. A relatively steep spectral index of $\alpha = -0.63 \pm 0.14$ was calculated using flux density measurements at component A, in agreement with the measurement made by \cite{2015MNRAS.454..287J}.

  \begin{table}
	\centering
    
	\begin{tabular}{ccc}
	    \hline 
        \hline \noalign{\smallskip}
	   Component& 4.32~GHz& 5.12~GHz \\
       		\hline \noalign{\smallskip}
	   A &  $93.7\pm 18.8$ &$82.0\pm 12.7$  \\
       B & $91.3\pm 18.8$  &  $80.0\pm 12.7$\\
       C1 & $46.9\pm 18.8$  &$42.1\pm 12.7$   \\
       C2 & $43.7\pm 18.8$  &$46.0\pm 12.7$   \\
       D &   $43.1\pm 18.8$ &$53.5 \pm 12.7$ \\
		\noalign{\smallskip}\hline
        \hline
	\end{tabular}
    	\caption{Radio flux density measurements made in \textmu Jy of the components of HS~0810+2554 observed with the e-MERLIN at 4.32~GHz and 5.12~GHz. Errors are derived from {\sc aips} task {\sc jmfit} and do not include an uncertainty on the setting of the overall flux density scale of 5-10\%.}
        \label{emerlintargetfluxes}
\end{table}

\subsection{All radio photometric data}
\label{photo}

\begin{figure}
  \centering
      \includegraphics[trim={1cm 0cm 2cm 1cm},clip,width=1\linewidth]{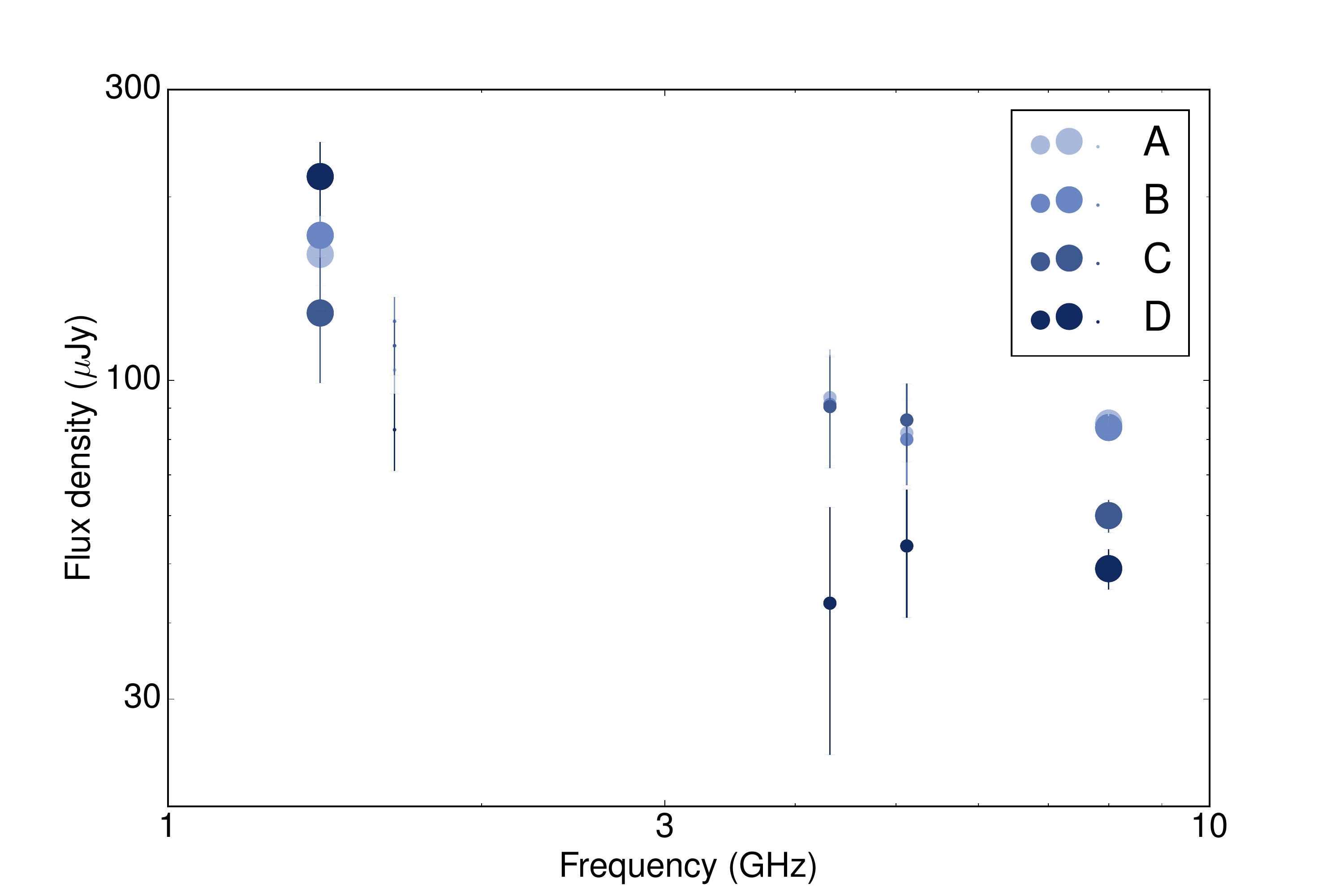} 
  \caption{Radio spectral energy distribution of HS~0810+2554, constructed using our EVN 1.65~GHz and e-MERLIN 4.32~GHz and 5.12~GHz observations, and the VLA 8~GHz and e-MERLIN 1.4~GHz observations made by \protect\cite{2015MNRAS.454..287J}. The size of the circles is proportional to the beamsize used in each observation. }
 \label{allf}
\end{figure}

Fig.~\ref{allf} presents all recorded flux density measurements for HS~0810+2554. Since interferometry is sensitive to the spatial scale determined by the size of the beam, any loss of flux density with an increase in resolution implies that at least some of the source occupies an angular extent up to and including the larger beamsize.  By accounting for the spectral index obtained from our e-MERLIN observations, we find a flux density loss of 37 ($\pm4$)\% with a beamsize reduction from 300 $\times$ 240 mas at 1.4~GHz to 12.0 $\times$ 8.5 mas at 1.65~GHz, when using the summed flux density from all components in each case. While most of the signal is contained within the very compact extent of the VLBI beam, some signal appears to originate from more diffuse structure.

Our analysis does not account for any intrinsic variation of the source flux density or position. Particularly on the timescales between the observations by  \cite{2015MNRAS.454..287J} and our own observations, quasar radio emission resulting from AGN activity has been observed to vary. Variation has even been detected within the few months difference in arrival times between components of this lens system \citep{2003ApJ...591L.103B,2018arXiv180105650B}. Additionally, millilensing by dark matter subhaloes can play a role, resulting in flux density variation over time if placed in front of a moving jet.  For example, subhalo masses of $10^6$ yield Einstein radii of $\sim$1 mas \citep{1998MNRAS.295..587M}, which could be traversed by a distant relativistic jet within a timescale of just $\sim$1 year in a typical quad lens system with modest magnification of the source plane. Either or both of these effects could be responsible for the variation in flux ratios of the more separated components from epoch to epoch, and could also contribute to the apparent spectral index.

% if subhaloes occupy an extent of $\sim$0.0003 arcseconds in the galaxy plane, as predicted by found by \cite{2017hsa9.conf..297V}

\section{Lens modelling}
\label{model}

\subsection{Methodology}

After revealing a clear detection of HS~0810+2554 at VLBI resolution, the EVN map was used to model the lens mass distribution and reconstruct the background source. Modelling software was written in {\sc python}. The script models the lens as an single isothermal ellipse (SIE) plus external shear and performs inverse ray tracing, populating the image plane with surface brightness values from the source plane. This is achieved by evaluating the deflection by the lensing mass at each point in the image plane and assuming a thin lens approximation. The source itself was modelled as a set of two two-dimensional Gaussian components. 

In total, a set, $\Theta$, of 11 free parameters were used to produce the model: source positions SX1, SY1 and SX2, SY2; lensing galaxy position and major axis, GX, GY and Gb; lensing galaxy ellipticity and position angle, GE and GPA; and external shear magnitude and position angle, GSM and GSA. The position of the lensing galaxy could not be fixed since it was undetected in the EVN image. Due to the point morphology of the lensed images, the ellipticity of the source components was set to zero, and their sizes to a nominal circular value. Since there is a degeneracy between the ellipticity of the SIE and the external shear, two approaches were attempted. First, the best model which used all 11 parameters was obtained. This model showed bi-modal distributions for both the lensing galaxy ellipticity and the shear magnitude, confirming the anticipated degeneracy and resulting in artificially large uncertainty values for these parameters. In order to determine these parameters and their  uncertainties in the absence of degeneracy, another model was obtained by first finding the best model without external shear and then fixing the SIE mass and ellipticity to leave only external shear and position angle to vary, as suggested by  \citep{1997ApJ...482..604K}. This model resulted in a better fit to observations.

%**use of a second same galaxy to change profile

Due to the point source nature of the lensed source, the objective function was constructed to measure a $\chi^2$ based on the positions of the eight lensed image components, driving the optimiser to find the model which would produce the minimum distances between the positions ${\bf x}^{\rm obs}$ of the observed components and positions ${\bf x}(\Theta)$ of the model components, and penalising outliers:

\begin{equation}
    \chi^2 = \sum_{i}^N\frac{|{\bf x}^{\rm obs}_i-{\bf x}_i(\Theta)|^2}{\sigma_i^2},
\end{equation}

where N is the number of image components to be used to constrain the model and $\sigma$ is the error on the position of each component $i$. Due to the possibility of intrinsic source variation, flux density values were not used as constraints to the model. As found by \cite{garrett1994global} and \cite{rusin2002high}, the relative astrometry of the lensed components can constrain the components of the relative magnification tensors without needing to use any flux density information.  Modelling in the $u-v$ plane of the interferometric data was not attempted, since the EVN provides good $u-v$ coverage, despite the absence of two telescopes for half of the observing time. 

We performed another model fit using the optical relative astrometric positions from HST observations from the CASTLES database  \citep{1999AIPC..470..163K} as additional constraints, adding two further parameters, SX3 and SY3, to the model. This allowed us to reconstruct both the radio and optical components simultaneously in the source plane. Given that the absolute astrometric uncertainties available for HST data are larger than those obtained using the EVN, the absolute astrometric positions of the HST components were not used as a constraint to the model. Instead, the HST component positions were overlaid onto the EVN component positions in a `best guess' by eye. The optical source was then modelled by constructing a second SIE with a variable position, GX2 and GY2,  but with a mass distribution fixed to the same values as the first SIE. Optical lensed image positions constrained the position of the second SIE which was shifted -- along with the relative positions of the optical lensed image positions and the relative position of the background optical source -- to the same position as the first SIE once model convergence was reached. Due to the growing parameter space, for this model we only performed a fit without external shear. 

Bayesian inference, a probabilistic method that has been applied to several model-fitting problems in strong and weak lensing \citep{2006MNRAS.371..983S,2007NJPh....9..447J,2007MNRAS.382..315M}, has been used to derive the probability distribution function associated with the model parameters.  Sampling of the parameter space was performed using the {\sc multinest} \citep{2009MNRAS.398.1601F,2008MNRAS.384..449F,2013arXiv1306.2144F} method implemented in {\sc python} using the {\sc pymultinest} \citep{2014A&A...564A.125B} package. Uniform priors were used for galaxy and source positions, galaxy major axis and galaxy and shear position angles, while logarithmic priors were used for galaxy ellipticity and shear magnitude. In order to reduce computation time, image resolution was downsampled by a factor of 2. 
% (Fig.~\ref{0810uv})

%the map was downsampled in order to  reduce computation time...**reduce the errors by running non-downsampled pixel positions

\subsection{Results}

%The best-fit model was produced without adding any external shear to the distortion of the main galaxy.  This model permitted two additional degrees of freedom, Since extragalactic observation records reveal no nearby objects in the field local to the main lens, we prefer...

%non-isothermal model... note allowing profile to vary can mask perturbations since degenerate with substructure

\begin{figure}
  \centering
    
      \includegraphics[trim={2.2cm 0cm 3.5cm 0cm},clip,width=1\linewidth]{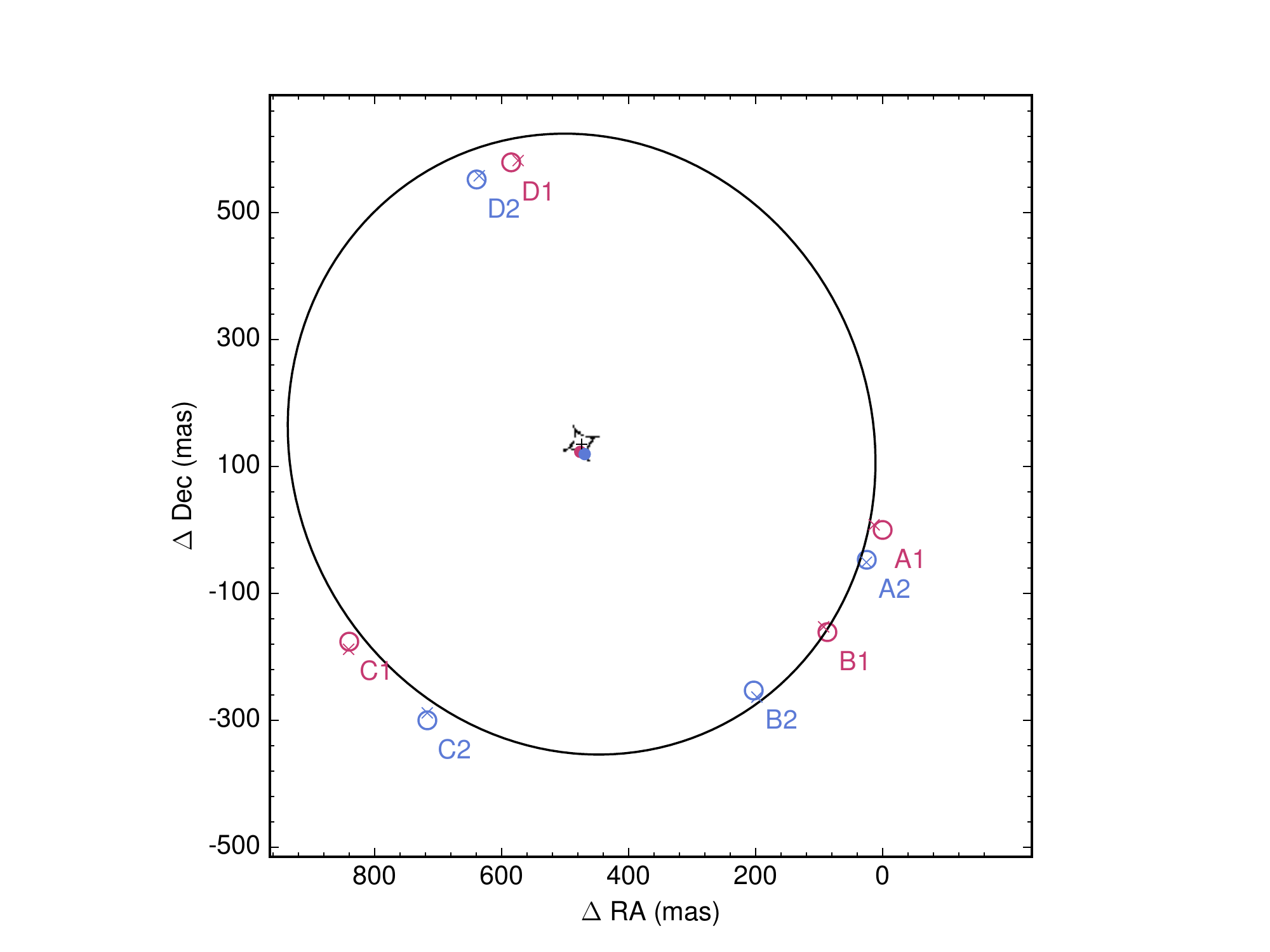}
  \caption{The best-fit model of HS~0810+2554 obtained using the positions of components in the EVN image as constraints. The lensing galaxy was modelled as an SIE and is centred on the vertical cross. The model includes a very small amount of external shear. The tangential critical curve is represented by the thin black curve and the caustic by the thick black curve. The predicted positions (crosses) of lensed components are plotted for comparison with observed component positions (open circles). The predicted positions of the components in the source plane (filled circles) are also plotted. Source 1 is represented by red markers and source 2 by blue. Model parameters and uncertainties are reported in Table~\ref{0810evnmodel} and derived parameters in Table~\ref{0810evnmodel2}. Model and observed component positions are reported in Table~\ref{0810evnmodel3}. All positions are plotted with respect to component A1.}
 \label{0810vlbi}
 \end{figure}

\begin{figure*}
\centering
\begin{adjustbox}{width=\linewidth}
\begin{tikzpicture}[      
        every node/.style={anchor=south west,inner sep=0pt},
        x=1mm, y=1mm,
      ]   
     \node (fig1) at (0,0)
       {\includegraphics[trim={11cm 5cm 15.5cm 11cm},clip,width=1\linewidth]{images/0810_1_margrevevn.pdf}};
     \node (fig2) at (120,135)
       {\includegraphics[scale=0.17]{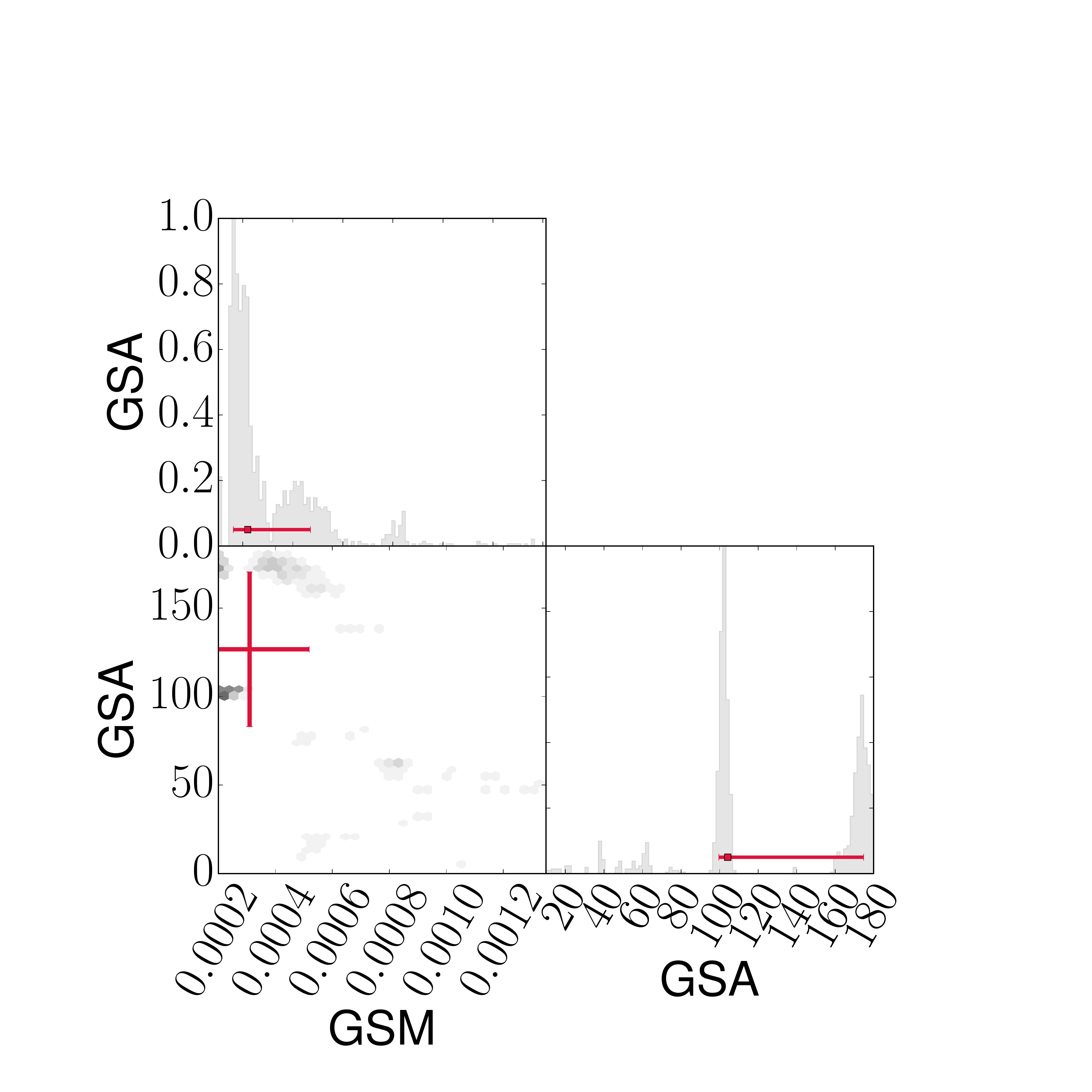}};  
\end{tikzpicture}
\end{adjustbox}
  \caption{Posterior probability distribution function (PDF) plots for the parameters of the best-fit model of HS~0810+2554. In each plot the error-bars are centred on the mean and extend to the 3$\sigma$ uncertainty values in the two-dimensional plots and to 1$\sigma$ in the one-dimensional plots. This model was obtained after first fitting for source positions and SIE mass model and position only, before then fixing the best-fit values of these parameters and allowing only external shear magnitude and angle to varying. The PDF plots are therefore separated into these two sets of parameters. This procedure avoided a degeneracy between lens ellipticity and external shear.}
 \label{marg}
\end{figure*}

The parameter values from the best-fit model using the EVN positional constraints only are presented in Table~\ref{0810evnmodel} and plots of the marginalised probability densities in Fig.~\ref{marg}. Derived values are presented in Table~\ref{0810evnmodel2}. Predicted lensed image positions are compared with observed positions in Table~\ref{0810evnmodel3} and plotted in Fig~\ref{0810vlbi}, along with source and galaxy positions and the extent of the tangential critical and caustic curves. Using positional errors from Table~\ref{0810evnmodel3}, the model represents an imperfect fit, suggesting astrometric perturbations arising from substructure along the line-of-sight. A fit of $\chi^2_{\rm red} = 1$ is only achieved by relaxing the RA and Dec positional constraints of the EVN image to within $\pm 8$ mas for the model containing a fixed external shear magnitude of zero, or to within $\pm 9$ mas when external shear is allowed to vary. The poorer fit after the addition of shear reflects the reduction in degrees of freedom combined with the low value of shear predicted. More detailed mass modelling of astrometric anomalies arising from possible substructures along the line-of-sight will be presented in a forthcoming paper.

 \begin{figure}
  \centering
      \includegraphics[trim={2.3cm 0cm 2.5cm 0cm},clip,width=1\linewidth]{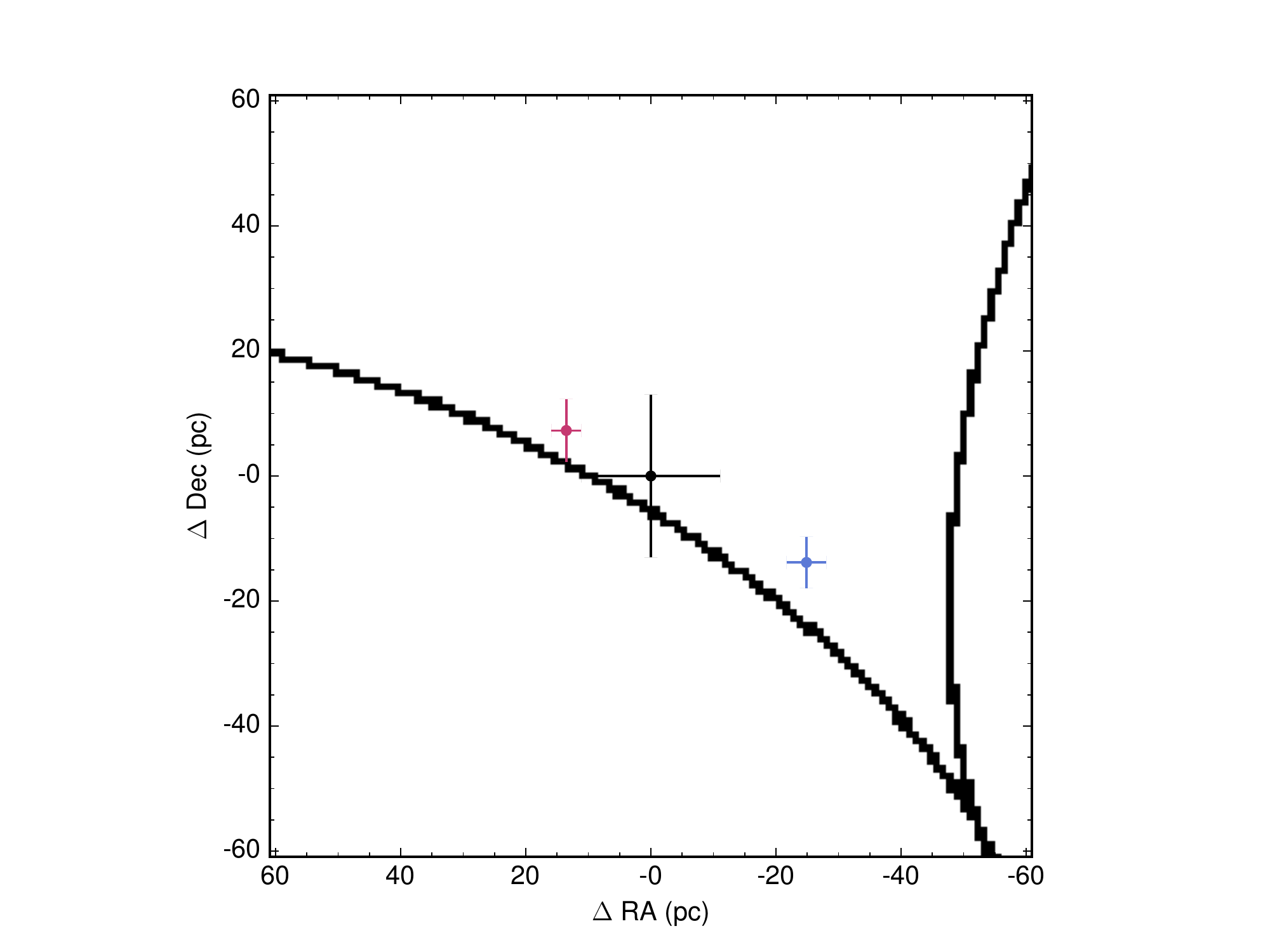} 
  \caption{Reconstruction of the lensed sources within the tangential caustic curve using both HST (black) positions and VLBI (blue and red)  positions from the best-fit EVN+HST model. Uncertainties (listed in Table~\ref{0810robust5}) were obtained from the {\sc multinest} Bayesian sampling tool and are represented at $1\sigma$ by the error bars. The components are plotted in the source plane at $z=1.51$, assuming a standard flat cosmology with $\Omega_{\rm m} = 0.27$ and $H_0 = 68\;{\rm km\;s^{-1}\;Mpc^{-1}}$. }
 \label{hst}
\end{figure}
 
  \begin{table*}
	\centering
	\begin{tabular}{clcccccc} % four columns, alignment for each
    	\hline
        	\hline\noalign{\smallskip}
                  
	     &&\multicolumn{3}{c}{EVN only} & \multicolumn{3}{c}{EVN+HST} \\
	    Parameter code&Parameter name & Mean & Sigma & Best-fit & Mean & Sigma & Best-fit\\
		\hline \noalign{\smallskip}
	%	\\
	    GX&Galaxy RA/mas & 474.8&2.3&474.0    &474.9&0.5&474.6\\
	    GY&Galaxy Dec/mas &142.9 &5.93& 135.2 &137.72&1.4&138.6\\
		SX1&Source 1 RA /mas & 477.6&2.4&476.4&477.4&0.5&477.3\\
		SY1&Source 1 Dec/mas &124.8&3.2& 123.0&122.6&1.2&123.42\\
		SX2&Source 2 RA /mas &467.2&2.5& 469.2&468.7&0.7&468.2\\
        SY2&Source 2 Dec /mas & 119.4&2.8& 119.0&117.9&0.9&118.6\\	
        GB&Einstein radius along major axis/mas & 477.6 &2.3& 475.4&479.1&0.7& 479.5\\
        GE&Galaxy ellipticity & 0.110 &0.020 & 0.077&0.094&0.0018&0.094\\
        GPA&Galaxy PA/$^{\circ}  $ &23.4& 0.3& 23.2&23.3&0.1&23.3\\
        GSM&External shear & 0.0003 &0.0002&0.0001 &--&--&-- \\
        GSA&External shear PA /$^{\circ} $&126.7 & 43.6 & 29.0&--&--&--\\
        &&&&&&&\\
        GX2&HST galaxy RA/mas & --&--&--    &451.0&1.8&450.5\\
	    GY1&HST galaxy Dec/mas & --&--&--  &145.0&1.9&145.0\\
		SX3&HST source RA /mas &-- &--&--&450.2&1.7&449.8\\
		SY3&HST source  Dec/mas &--&--&--&128.1&1.9&128.1\\

    %    \\
		\noalign{\smallskip}\hline
        	\hline \noalign{\smallskip}
	\end{tabular}
    	\caption{The best-fit, mean, and uncertainty on the mean are presented for the parameters of the maximum-likelihood models of HS~0810+2554 obtained using positional constraints from the EVN image only and the EVN image plus the HST positions from the CASTLES database \citep{1999AIPC..470..163K}. The fitting was performed using the {\sc multinest } Bayesian sampling tool. Galaxy and source positions are reported with respect to lensed imaged A1. Since absolute astrometric uncertainty associated with the HST image positions is larger than the absolute astrometric uncertainty associated with the EVN positions, HST absolute astrometry was not used as a constraint. Instead, the whole HST system was initially placed at a nominal `best guess' position with respect to the EVN system. A model was then constructed to contain two, identical, SIEs, one for each system. After convergence, the whole HST system was shifted by the difference between the best fit positions of the two lensing galaxies, i.e. ${\rm GX-GX2} =24.1$ and ${\rm GY-GY2}=-6.4$ in order to align the HST SIE with the EVN SIE, and to determine the respective locations of radio and optical components in the source plane (Fig.~\ref{hst}). The positions of the HST galaxy and source are quoted before performing this shift, in order to report the respective uncertainties. }
        \label{0810evnmodel}
\end{table*}

\begin{table}
	\centering
	\begin{tabular}{lc} % four columns, alignment for each
    	\hline 
		\hline \noalign{\smallskip}
        Density slope (2$\equiv$ isothermal) &  $\equiv$2.0\\
        Source separation /mas &8.2 $\pm$ 6.2 \\
        Source position angle /$^{\circ}$   & 60.9 $\pm$ 23 \\
        Source centroid position /mas &$2.8\pm 3.1$ W, $14.2\pm 6.1$ S \\
            %    \\
		\noalign{\smallskip}\hline
        	\hline 
	\end{tabular}
    	\caption{Additional properties of the system. The source properties are derived from the best-fit parameters of Table~\ref{0810evnmodel} for the model fit using EVN positional constraints only. The source centroid position is quoted with respect to the galaxy position.}
        \label{0810evnmodel2}
\end{table}

\begin{table*}
	\centering
	\begin{tabular}{ccccccc} % four columns, alignment for each
     	\hline 
        	\hline \noalign{\smallskip}
	%	\\
    	   &\multicolumn{2}{c}{Observed}&\multicolumn{4}{c}{Modelled}\\
   && &\multicolumn{2}{c}{EVN only}&\multicolumn{2}{c}{EVN+HST}\\

		Comp. & $\Delta$RA (mas) &  $\Delta$Dec (mas) &$\Delta$RA (mas) &$\Delta$Dec (mas)&$\Delta$RA (mas) &  $\Delta$Dec (mas) \\
        		\hline \noalign{\smallskip}
	    A1 & $0.0\pm 1.4$ &  $0.0 \pm 1.0$ &$ 13.5$ &  $8.1$ &9.9&8.2\\
	    A2 & $25.8\pm 1.2$ &  $-47.2 \pm 1.5$ &$25.6$ &  $ -50.9$ &22.5&-57.0\\
	    B1 & $87.5\pm 1.2$ &  $-161.8 \pm 0.8$ &$93.0$ &  $-152.6$ &94.5&-158.4\\
	    B2 & $203.4\pm 1.0$ &  $-253.4\pm 1.0$ &$198.0$ &  $-263.4$ &197.0&-267.4\\
	    C1 & $840.5\pm 1.6$ &  $-176.3\pm 1.2$ &$841.2 $ &  $-187.6$&844.0&-194.0\\
	    C2 & $717.9\pm 1.5$ &  $-300.1 \pm 0.9$ &$717.2 $ &  $-288.2$&716.2&-293.9 \\
	    D1 & $585.1\pm 1.3$ &  $579.9\pm 1.3$ &$574.0$ &  $ 582.0 $ &578.0& 584.0\\
        D2 & $640.6\pm 4.1$ &  $552.2 \pm 2.1$ &$635.3 $ &  $558.0$ &637.4&558.3\\

        A HST & $-6.0\pm5$&$-4\pm5$&--&--&-6.4&-20.8\\
        B HST &$87.0\pm5$ &$-167.0\pm5$&--&--&84.9&-174.9\\
        C HST & $794\pm5$&$-261\pm5$&--&--&780.1&-225.9\\
        D HST &$604\pm5$ &$575\pm5$&--&--&578.0&582.0\\
        
    %    \\
		\noalign{\smallskip}\hline
        	\hline 
	\end{tabular}
    	\caption{Observed and modelled positions for each lensed component of HS~0810+2554. All positions are measured with respect to component A1 in the EVN image. The observed HST positions were set to a nominal reference position prior to model fitting, so that the difference between the between EVN lensing galaxy position, GX and GY, and HST lensing galaxy position, GX2 and GY2, was effectively a free parameter in the fit. After model convergence the quoted observed and predicted positions were then shifted by this difference, ${\rm GX-GX2} =24.1$ and ${\rm GY-GY2}=-6.4$ (using the best-fit positions from Table~\ref{0810evnmodel}), in order to view the relative physical positions of all components in Fig.~\ref{0810vlbihst}.}
        \label{0810evnmodel3}
\end{table*}

The parameters values of the best-fit model are in reasonable agreement with the model produced by \cite{2015MNRAS.454..287J} (hereafter J15), the values of which were not used as priors or constraints to the VLBI model. The galaxy critical radius, the source centroid position and the source position angle -- defined as the angle of orientation between the background source components -- all agree to within $1\sigma$.  Galaxy ellipticity and external shear do not agree; while our model predicts a minimal external shear value and significant galaxy ellipticity, the model of J15 predicts the opposite. This can be explained by the degeneracy between the two forms of lensing distortion \cite{koch2004}. This was borne out during the modelling process, which found complementary, bi-modal, distributions in the marginal plots of galaxy ellipticity and external shear when both parameters were searched over simultaneously. The shear position angle of J15 agrees to just outside $1\sigma$ of the galaxy position angle of our model. The low shear in our model could therefore be explained by the degeneracy between ellipticity and shear, making it difficult to disentangle intrinsic and extrinsic shear when the two directions are almost identical.  

%We note that intrinsic perturbations can be partially masked by degeneracies between substructure and parameters used in the fit. Allowing galaxy position, mass slope and external shear to vary can mask intrinsic perturbations by an order of magnitude. \cite{1997ApJ...482..604K}, for example, have found that while galaxies modelled as SIEs rarely fit the data well, the use of large external shear amplitudes tend to result in good fits, which suggests that significant unknowns become masked by adding additional model parameters. We will therefore present results from using both EVN and HST constraints, using a fixed galaxy position and accounting explicitly for shear contribution from a galaxy group located 58 arcsec west of HS~0810+2554 \citep{2011ApJ...742...93A}. 

Lensed image flux density values obtained from the model predict ratios greater than 10:1 between images of both sources forming in the general regions of A or B and images forming in the region of C. This prediction is discrepant with the observed flux density values, which find comparable values between images A, B and C for both sources (Table~\ref{0810robust5}).  The apparently boosted flux density in region C could be the result of millilensing by dark matter substructure. However, the evidence from radio observations (Fig.~\ref{allf}) taken at other epochs shows that the relative flux density values between all regions have varied over timescales greater than one year, implying that the cause of the discrepancy is due to intrinsic variation of the source: either the brightening and dimming of AGN activity, or the motion of jets across the plane of the lensing caustic.

\begin{figure}
  \centering
    
      \includegraphics[trim={2.2cm 0cm 3.5cm 0cm},clip,width=1\linewidth]{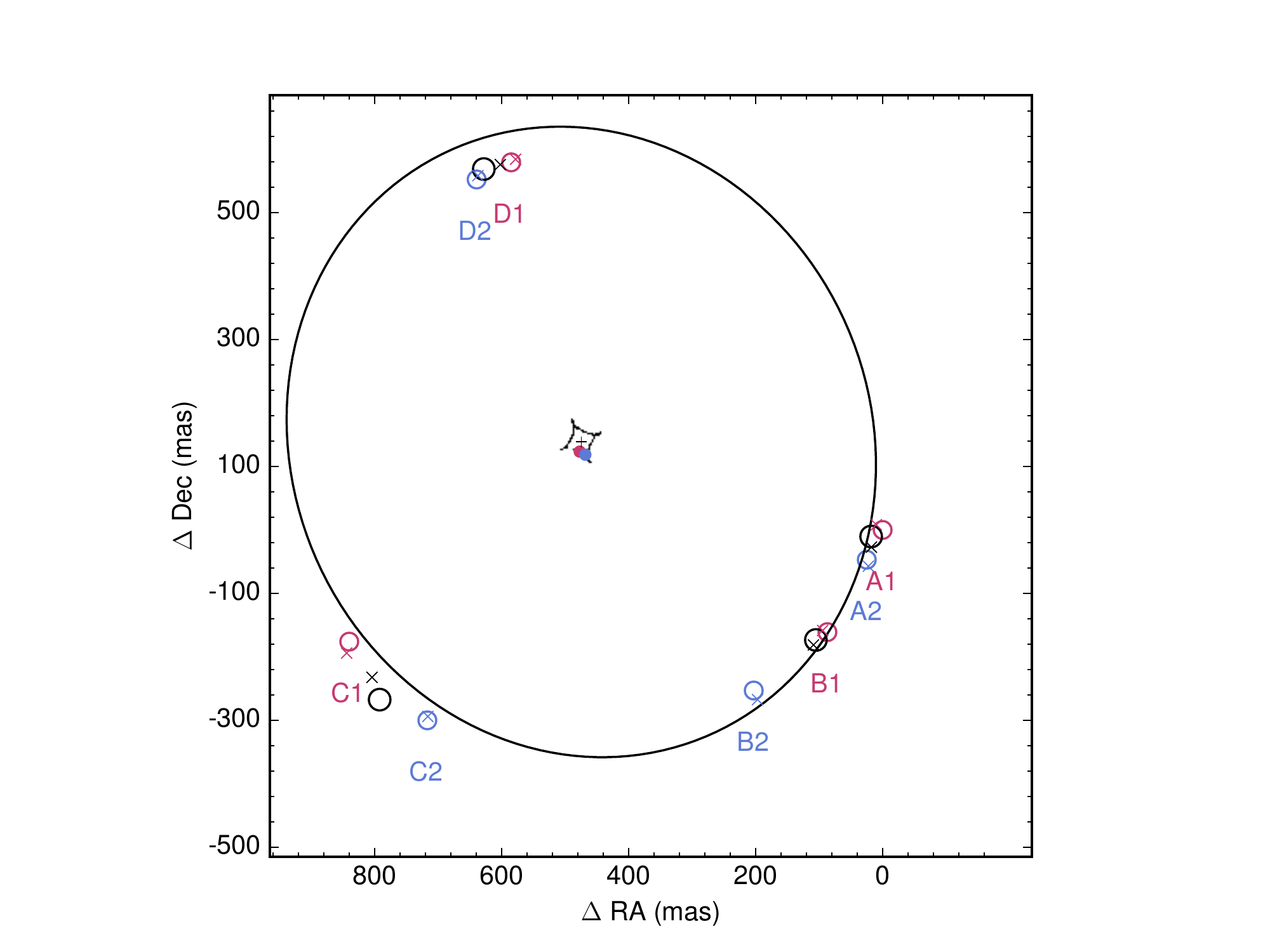}
  \caption{The best-fit model of HS~0810+2554 obtained using the positions of components in the EVN image and HST observations as constraints. The lensing galaxy was modelled as an SIE and is centred on the vertical cross. The tangential critical curve is represented by the thin black curve and the caustic by the thick black curve. The predicted positions (crosses) of lensed components are plotted for comparison with observed component positions (open circles). The predicted positions of the components in the source plane (filled circles) are also plotted. Source 1 is represented by red markers,  source 2 by blue and the source observed by HST in black. Model parameters and uncertainties are reported in Table~\ref{0810evnmodel} and derived parameters in Table~\ref{0810evnmodel2}. Model and observed component positions are reported in Table~\ref{0810evnmodel3}. All positions are plotted with respect to component A1.}
 \label{0810vlbihst}
 \end{figure}
 
 \begin{figure*}
  \centering
      \includegraphics[trim={20cm 60cm 58cm 10cm},clip,width=1\linewidth]{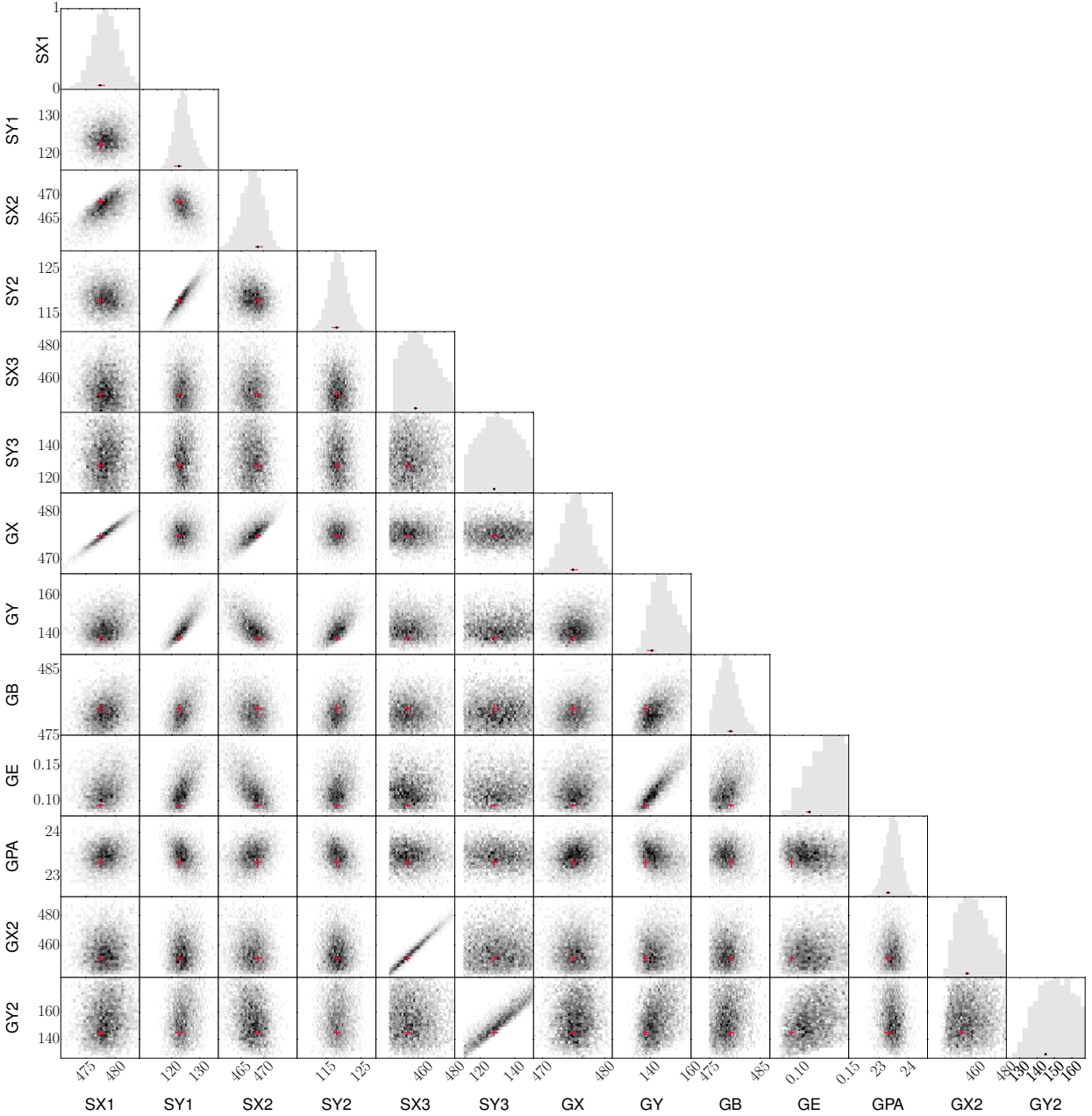} 
  \caption{Posterior probability distribution function (PDF) plots for the parameters of the best-fit model of HS~0810+2554 using positional constraints from the EVN image and HST observations. In each plot the error-bars are centred on the mean and extend to the 3$\sigma$ uncertainty values in the two-dimensional plots and to 1$\sigma$ in the one-dimensional plots. }
 \label{hstpdf}
\end{figure*}

Tables~\ref{0810evnmodel} and ~\ref{0810evnmodel3} report parameter estimates and predicted lensed image positions obtained by using both EVN and HST observations. Fig~\ref{hstpdf} presents associated marginalised distribution plots. The tangential caustic and critical curves and the predicted and observed positions are plotted in Fig.~\ref{0810vlbihst}. Adding the HST constraints allows us to predict a source configuration featuring the two radio components linearly aligned on either side of the optical quasar core (Fig.~\ref{hst}). Inclusion of HST positions also allows us to compare the separation of between our galaxy position and HST component A with the separation between observed HST galaxy position and observed component A. With respect to HST image A, RA and Dec positions of the galaxy are predicted to be 456.5 $\pm$1.8~mas and 149.0 $\pm$1.9~mas, respectively,  agreeing to within 1$\sigma$ of the values reported in the CASTLES database.

%This is understandable in terms of a degeneracy between galaxy position with respect to the source - which would prefer a location near the tangential caustic in order to reproduce the fold configuration of this quad lens system - and the galaxy ellipticity, which directly defines the caustic. Such a degeneracy is indeed apparent in the two-dimensional marginalised plots (Fig.~\ref{marg}). Therefore, it appears that the main discrepancy between the two models is in the form of additional ellipticity of the lensing mass in the VLBI model.

Magnification factors associated with each lensed image component were obtained by mapping the determinant of the lensing Jacobian used to perform ray-tracing during the modelling process. Magnification is very strong in the region of images A and B, as would be expected in a merging fold system, and factors of 70, 46, 67 and 54 at the locations of A1, A2, B1 and B2, respectively, are found. These values are in fair agreement with a value of 50 from models of the narrow-line region made by \cite{2011ApJ...742...93A}, and of 40 from models of X-ray emission made by \cite{2016ApJ...824...53C}, given the different locations of the various components in the high magnification regime next to the very small astroid caustic. Using the upper limit to the deconvolved source size of 1.55 mas from the EVN map, and assuming a standard flat cosmology with $\Omega_{\rm m} = 0.27$ and $H_0 = 68\;{\rm km\;s^{-1}\;Mpc^{-1}}$, a conservative magnification factor of 50 gives an upper limit to the intrinsic component size in the source plane of just 0.27 pc. Furthermore, unresolved VLBI flux density measurements from the lensed plane at A1, A2, B1 and B2 imply an intrinsic source flux density no higher than 880~nJy at the source plane: the faintest radio source ever to be observed. 

%edge on galaxy causing anomaly: not likely here accordingf to evidence from HST which indicates an elliptical red gal \cite{813facf8c87c4db3af32dc5ee75d2067}

\section{Discussion and conclusions}

The EVN observation of 0810+2554 provides the highest resolution image ever made of radio activity within a RQQ. The final map shows eight lensed components in total, and modelling of the system finds that the background source contains two highly compact components -- one itself a double component -- separated by $\sim$100~pc and positioned on the small astroid caustic. Intensity measurements find that only 37 ($\pm4$)\% of flux density is lost as resolution is increased from the 220~mas scale to the 5~mas scale. Thanks to a resulting lensing magnification factor of  $\sim$50, source structure can be seen on the sub-parsec scale, at a finer detail even than the closest unlensed quasars. 3C273, for example, at a redshift of $z= 0.15$, can be resolved to a 3~pc scale at a resolution of 1~mas. The resolution of this observation finds each unresolved individual component to be no larger than 0.27~pc in diameter.  The magnification factors also imply an intrinsic source flux density of just 880~nJy, making this object the faintest ever observed radio source and the first from the nJy regime.

The EVN map is complemented by observations using the e-MERLIN array at C~band. Maps made at two different frequencies within the band allowed a relatively steep spectral index of $\alpha = -0.63 \pm 0.14$ to be measured, in agreement with the value found by \cite{2015MNRAS.454..287J}. Further use of the unique combination of strong lensing with VLBI resolution and sensitivity allowed strong constraints to be place on the brightness temperature of the radio components. Several authors, including \citep{1991ApJ...378...65C,1992ARA&A..30..575C,1994MNRAS.266..455M}, place an upper limit on the brightness temperature of starburst regions, at $10^{4-5}$~K, although \cite{1991ApJ...378...65C} based this limit on observations made with a 0.25 arcsec resolution and therefore could not measure the brightness temperatures of sub-pc scale sources. With brightness temperature measurements that exceed $10^6$~K, it appears very likely that the radio emission seen in HS~0810+2554 originates from AGN jet/core activity.

One alternative hypothesis is the possibility that the observed emission results from radio supernovae, which have been observed to display brightness temperatures in excess of $10^6$~K and are very compact \citep{1991ApJ...378...65C,1998ApJ...493L..17S,2001ASPC..249..119C}. However, such sources vary over timescales of $\sim$1 year \citep{2015MNRAS.446.3687P,2006evn..confE..51B}, both in flux density and in angular extent, and individual supernovae are likely to become resolved out by VLBI baselines over the timescale since HS~0810+2554 was first observed. 

It is possible that a high rate of supernova turnover in a starburst region could result in a compact area of consistently high brightness temperature. However, such regions generally occupy over 100--200~pc in extent, and the resolution of this observation would have revealed multiple such sources in the background plane, if present. For example, observations of strong nuclear activity within Arp 220 by \cite{1998ApJ...493L..17S} found 12 radio supernovae within a 75 $\times$ 75 pc region, and \cite{1994MNRAS.266..455M} found 40 supernova remnants distributed across the inner 700 pc of starburst galaxy M82. Compact AGN jets components, on the other hand, extend in a generally linear formation, occupying regions smaller than 0.1~pc near the quasar core \citep{1999Natur.401..891J}.

In addition, following arguments also made by \cite{1995MNRAS.274L...9B} in the study of closer, unlensed RQQ, the maximum supernova luminosity at radio frequencies is of order $\sim10^{21}\;{\rm W\;Hz^{-1}}$ at 1.6~GHz \citep{1987AJ.....94...61R}. Assuming a magnification factor of 50, the intrinsic flux density of HS~0810+2554 of 880 nJy yields a rest frame luminosity of $7.5\times10^{21}\;{\rm W\;Hz^{-1}}$ at redshift $z = 1.51$ using the derived spectral index of $\alpha = -0.63$. Therefore, tens of supernovae would be required to exist in three separate regions each smaller than the extent of the sub-parsec source component scale, in order for the observed flux density levels to be seen. 

Our final piece of evidence comes from the inclusion of HST data as constraints to the lens model. This allowed the source reconstruction of three components -- two radio and one optical -- to be performed. The results show a clear linear alignment of the three background sources, with two opposing jet components situated very close to the quasar core. This presentation suggests the morphology of a compact symmetric object (CSO), in which compact jets are located on the parsec-scale either side of the core \citep{1994ApJ...432L..87W,csos}. Observations of the rapid advancement of hotspots in several CSOs suggest that CSOs represent a population of very young radio loud sources, with ages under $\sim$1000 years \citep{1996ApJ...460..634R,1998A&A...337...69O}. Further observations have shown, however, that although some CSOs may evolve according to self-similar models into far larger and more luminous FR II objects, a large fraction are either short-lived \citep{2012ApJ...760...77A} or produce recurrent small-scale jets (e.g. \cite{1998A&A...336L..37O}). The jets observed in HS~0810+2554 may therefore represent either the early stage of jet production within the AGN core or an intermittent type of lower-power jet activity. It is important to note that, due to the finite scale of the strong lensing magnification region, our observations do not extend to the kpc scales typically associated with larger jet and lobe structure. We therefore cannot rule out the presence of such emission in this case.

%This finding  strengthens the evidence found by \cite{1995MNRAS.274L...9B} for a connection between FR I AGN and RQQ and suggests a simplified version of the unification model based on orientation \citep{1987slrs.work..104S}, incorporating RQQ as edge-on projections of BL Lac parents which would be consistent with the association between BL Lacs and FR I sources noted by e.g. \cite{1983MNRAS.204P..23B,1984ApJ...279...93W}.

%\cite{2018arXiv181010245L} find spectral index correlation with L over Ledd for RQQ - but RLQ spec determined by MBH

By finding that even a quasar with the very faintest radio emission can contain jet activity, and that the radio emission from this activity can dominate the total radio output, this work provides direct evidence for a continuum model of jet production in the quasar population. It extends the work of \cite{1538-4357-468-2-L91,1995MNRAS.274L...9B,blundell1998central}, who have found similar evidence in sources of intermediate radio flux density. By imaging the radio emission from a quasar of monochromatic radio luminosity $L_{1.6\;{\rm GHz}}\sim 10^{22}\;{\rm W\;Hz^{-1}}$, it also suggests that the source count upturn at the faint end of the quasar luminosity function may not represent a population of different -- star-forming -- sources as suggested by \cite{0004-637X-768-1-37} and \cite{2016arXiv160804586K}. However, a investigation by \cite{2008arXiv0811.3421W} found that a VLA and e-MERLIN detection of RQQ RX~J1131$-$1231 was not replicated using VLBI, implying the absence of jets in this source. Larger studies of RQQs at high resolution will be required in order to determine the full contribution of jet-powered RQQs to the quasar luminosity function. 

%We suggest that the apparent RQQ/RLQ dichotomy results from the same origin as the FR I/FR II dichotomy based on AGN radio luminosities: AGN jet production is occurring in each case, but is modified either by external factors such as environment or, intriguingly, by internal factors such as black hole spin \citep{2010ASPC..427..357S,zirbel1995fr,2010ApJ...711...50T}.

%new one arxiv rqq

It is possible that jet production is able to coexist within RQQ alongside star-forming activity, with both phenomena triggered by the violent inflow of matter towards the SMBH. \cite{2018MNRAS.476.5075S}, for instance, studied a sample of gravitationally lensed quasars and found high levels of star-formation in general, concluding that a coexistence of dust-obscured star formation and AGN activity is typical of most quasars. Several authors, e.g.  \cite{KEWLEY19991045,doi:10.1093/mnras/stu234} have found  starburst and AGN mixing in lower luminosity sources. These findings, along with the observations of HS~0810+2554,  all suggest that jet activity does not represent a transitional stage of AGN activity but can feature at any stage of the AGN life-cycle.

Of further significance is that the FIR luminosity measurements made by \cite{2018MNRAS.476.5075S} found that HS~0810+2554 does not display the expected radio excess from the radio--FIR correlation typically associated with AGN activity.  While we have found clear evidence for jet-dominated radio emission, observations of X-ray absorbing gas by \cite{2014ApJ...783...57C,2016ApJ...824...53C} within HS~0810+2554 found evidence for a relativistic gas outflow. Since such ultrafast winds must originate from the active nucleus \citep{2012ARA&A..50..455F}, since AGN winds have been shown to contain heated dust \citep{2018MNRAS.tmp.2800B}, and since the high dust temperature of this source is consistent with dust heated by the AGN \citep{weiss}, we suggest that the location of this source on the correlation may arise due to the coexistence of kinetic and radiative modes of AGN feedback.  The radio--FIR correlation could, therefore, conceal emission processes other than star-formation activity, and may not be a reliable method of distinguishing jetted from non-jetted quasars. The coexistence of jet activity and AGN winds also has implications for the understanding of AGN feedback modes within RQQ, for which radiation pressure is typically invoked over mechanical effects as the driving feedback mechanism \cite{2018NatAs...2..181W}.

RQQ PG~1115+080 has also been observed by \cite{2018MNRAS.476.5075S}. Again, a relatively high dust temperature has been measured, suggesting that the object lies, with HS~0810+2554, on the radio--FIR correlation of \cite{2010A&A...518L..31I}, yet X-ray observations by \cite{0004-637X-595-1-85} have again found evidence for a relativistic gas outflow. We have recently conducted EVN observations of this additional source and will be able to access the correlated data shortly. VLBI detection of this object would again allow us to identify any bright and compact emission typical of an AGN, ruling out starburst activity as the primary mechanism at play. Detection of another jetted source in this region would demand an urgent review of the use of the correlation to classify star-forming and AGN activity. 

Observations of HS~0810+2554 have also provided the opportunity to study the lensing galaxy itself. Modelling of the lens has suggested the presence of a non-smooth mass distribution in the vicinity of Einstein radius of image plane, producing astrometric perturbations of the compact VLBI lensed images. Our findings add to those of others (e.g. \citealt{2004MNRAS.350..949B,2018MNRAS.478.4816S}) who have been unable to fit smooth mass models to the data. We rule out the possibility of substantial baryonic disk structure since HST observations by \cite{813facf8c87c4db3af32dc5ee75d2067} find that the main lens appears to be an early type elliptical galaxy. Relaxing the model to include the radial profile of the lens may reduce the positional offsets between the observed and modelled lensed components. However, studies by \cite{0004-637X-659-1-52} and \cite{2003MNRAS.345.1351E} have found that using a simple smooth mass model rarely fits the data well. Instead, angular structure in the lens at different scales ranging from the SIE and down can reproduce observed offsets.  Additionally, \cite{1997ApJ...482..604K} have found that allowing the galaxy position, mass slope and external shear of the macromodel to vary can mask intrinsic perturbations by an order of magnitude. More quantitative constraints on the possible presence of the substructures along the line-of-sight as derived from the astrometric anomalies suggested by our model of HS~0810+2554 will be presented in a forthcoming paper.

%Further modelling analysis could be performed in order to investigate additional lens mass parameters. We note, however, that intrinsic perturbations are partially masked by degeneracies between substructure and parameters used in the fit \citep{1997ApJ...482..604K}. Allowing galaxy position, mass slope and external shear to vary can mask intrinsic perturbations by an order of magnitude.  \cite{1997ApJ...482..604K}, for example, have found that while galaxies modelled as SIEs rarely fit the data well, the use of large external shear amplitudes tend to result in good fits, which suggests that significant unknowns become masked by adding additional model parameters.

\section*{Acknowledgements}

PH acknowledges receipt of an STFC studentship. HVA acknowledges the Spanish Plan Nacional de Astronom\'ia y Astrofis\'ca under grant AYA2016-76682-C3-2-P. The European VLBI Network is a joint facility of independent European, African, Asian, and North American radio astronomy institutes. Scientific results from data presented in this publication are derived from the following EVN project codes: EJ016A and EJ016B. e-MERLIN is a National Facility operated by the University of Manchester at Jodrell Bank Observatory on behalf of STFC. We also thank James Nightingale for a helpful discussion on model optimisation.

%%%%%%%%%%%%%%%%%%%%%%%%%%%%%%%%%%%%%%%%%%%%%%%%%%



\bibliographystyle{mnras}
\bibliography{phd,0810} % if your bibtex file is called example.bib

%%%%%%%%%%%%%%%%%%%%%%%%%%%%%%%%%%%%%%%%%%%%%%%%%%

%%%%%%%%%%%%%%%%% APPENDICES %%%%%%%%%%%%%%%%%%%%%
\newpage
\appendix

\section{EVN 1.6~GHz map of HS~0810+2554}

A contour plot is provided in order to show the uncovered background.

 \begin{figure*}
  \centering
      \includegraphics[trim={0cm 0cm 0cm 0cm },clip,width=1\linewidth]{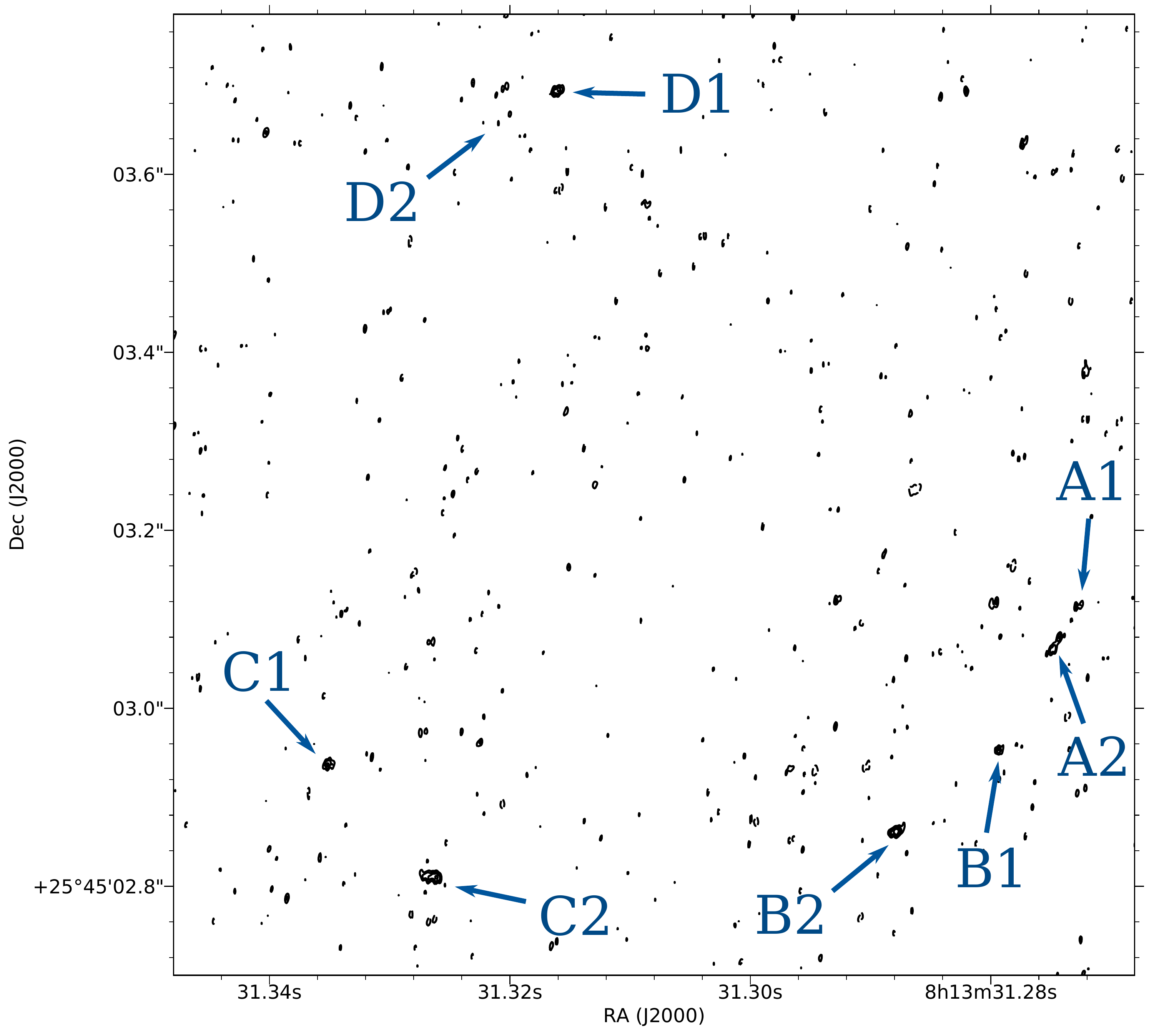}
  \caption{EVN image of HS~0810+2554 at 1.65~GHz produced using a natural weighting scheme.  Contours are drawn at (-3,3,4,5,6) $\times7.2$ \textmu Jy. The peak surface brightness is 52~\textmu Jy beam$^{-1}$, at component B2. The beam is at full width of half maximum (FWHM) at 12.0 $\times$ 8.5~mas at a position angle of -3$^{\circ}$.}
 \label{contall}
\end{figure*}

%%%%%%%%%%%%%%%%%%%%%%%%%%%%%%%%%%%%%%%%%%%%%%%%%%

% Don't change these lines
\bsp	% typesetting comment
\label{lastpage}
\end{document}